\documentclass[alpha-refs]{nbdt-article}

\usepackage{siunitx}
\usepackage{gensymb}
\usepackage{graphicx}
\usepackage{textcomp}

\papertype{Original Article}
\paperfield{Journal Section}

\title{Overlapping neural representations for the position of visible and imagined objects}

\author[1,2,3]{Amanda K. Robinson}
\author[1,2,3,4]{Tijl Grootswagers}
\author[1]{Sophia M. Shatek}
\author[1]{Jack Gerboni}
\author[1]{Alex Holcombe}
\author[1,2]{Thomas A. Carlson}

\affil[1]{School of Psychology, University of Sydney, NSW, 2006, Australia}
\affil[2]{Perception in Action Research Centre, Macquarie University, Sydney, NSW, 2109, Australia}
\affil[3]{Department of Cognitive Science, Macquarie University, NSW, 2109, Australia}
\affil[4]{The MARCS Institute for Brain, Behaviour and Development, Western Sydney University, Sydney, Australia}

\corraddress{Amanda K. Robinson, School of Psychology, University of Sydney, NSW, 2006, Australia}
\corremail{amanda.robinson@sydney.edu.au}

\fundinginfo{T.A.C: Australian Research Council Discovery Projects (DP160101300 and DP200101787); A.K.R: Australian Research Council Discovery Early Career Research Award (DE200101159)}

\runningauthor{Robinson et al.}

\begin{document}

\maketitle

\begin{abstract}
Humans can covertly track the position of an object, even if the object is temporarily occluded. What are the neural mechanisms underlying our capacity to track moving objects when there is no physical stimulus for the brain to track? One possibility is that the brain “fills-in” information about imagined objects using internally generated representations similar to those generated by feed-forward perceptual mechanisms. Alternatively, the brain might deploy a higher order mechanism, for example using an object tracking model that integrates visual signals and motion dynamics \citep{Kwon_2015} . In the present study, we used electroencephalography (EEG) and time-resolved multivariate pattern analyses to investigate the spatial processing of visible and imagined objects. Participants tracked an object that moved in discrete steps around fixation, occupying six consecutive locations. They were asked to imagine that the object continued on the same trajectory after it disappeared and move their attention to the corresponding positions. Time-resolved decoding of EEG data revealed that the location of the visible stimuli could be decoded shortly after image onset, consistent with early retinotopic visual processes. For processing of unseen/imagined positions, the patterns of neural activity resembled stimulus-driven mid-level visual processes, but were detected earlier than perceptual mechanisms, implicating an anticipatory and more variable tracking mechanism. Encoding models revealed that spatial representations were much weaker for imagined than visible stimuli. Monitoring the position of imagined objects thus utilises similar perceptual and attentional processes as monitoring objects that are actually present, but with different temporal dynamics. These results indicate that internally generated representations rely on top-down processes, and their timing is influenced by the predictability of the stimulus. All data and analysis code for this study are available at \url{https://osf.io/8v47t/}. 

\keywords{MVPA, imagery, visual perception, Electroencephalography}
\end{abstract}

\section{Introduction}
Internally-generated representations of the world, as opposed to stimulus-driven representations, are important for day-to-day tasks such as constructing a mental map to give a stranger directions, remembering where you last saw a lost item, or tracking the location of a car that becomes occluded by another vehicle. In these cases, there is little or no relevant perceptual input, yet the brain successfully constructs a picture of relevant visual features such as object form and spatial position. Internally-generated representations have been studied with tasks involving imagery, mental rotation, and perception of occluded objects. Such tasks seem to involve different cognitive strategies, yet their neural mechanisms have some similarities. It is clear that internally-generated representations rely on similar brain regions to stimulus-driven representations \citep{Lee_2012, Reddy_2010} but they appear to have different temporal dynamics \citep{Dijkstra_2018}, raising the question of how exactly these internal representations are formed.

Top-down processing appears to play an important role in generating internally representations. Current theories of mental imagery are based on similarities between perception and imagery, with a greater focus on bottom-up processing in perception and top-down processing in imagery (for review, see \citet{Pearson_2019}). Neuroimaging work has shown increases in brain activation within early visual cortical regions when participants engage in imagery, in a similar way to viewing the same stimuli \citep{Kosslyn_1993, Le_Bihan_1993}, but there is more perception-imagery overlap in higher level brain regions such as ventral temporal cortex \citep{Lee_2012, Reddy_2010}. Imagery involves greater flow of information from fronto-parietal to occipital regions than perception, indicating that top-down or feedback-like processes mediate internally generated representations \citep{Dentico_2014, Dijkstra_2017, Mechelli_2004}. During imagery, neural activation within the ventral stream is consistent with generative feedback models of information flow from higher-level to low-level visual regions \citep{Breedlove_2020}. Consistent with this account, recent work using magnetoencephalography and time-resolved decoding showed that imagery of faces and houses involves similar patterns of activation as viewing those stimuli, but with different temporal dynamics \citep{Dijkstra_2018}. In the Dijkstra et al. (\citeyear{Dijkstra_2018}) study, imagery-related processing was delayed and more diffuse than perception, which showed multiple distinct processing stages. A follow-up study suggested that the order of perceptual processes is reversed in imagery \citep{Dijkstra_2019}. Together, these results suggest that imagery uses at least some of the same mechanisms as perception but is initiated in higher-level brain regions rather than being driven by perceptual input.

Another mechanism originating in higher-level brain regions that might be intrinsically linked to internal representations is spatial attention. Directing attention to a location enhances processing of stimuli that appear there \citep{Posner_1980}. Reduced amplitude alpha-band ($\sim$10Hz) oscillations in visual cortex have been linked to covertly attending to a specific region in space \citep{Worden_2000}. Additionally, time-resolved decoding has found that attended locations could be decoded from the neural signal even before a stimulus appeared \citep{Goddard_2019}. It follows that spatial imagery tasks that require internal representations of objects with specific positions or orientations, such as in occlusion or mental rotation, might also inherently involve spatial attention. Indeed, alpha-band activity has been found to track spatial locations held in working memory \citep{Foster_2016}. Interestingly, a recent study found evidence that imagery and perception share neural processes in the alpha-band frequency linked to high-level visual processing, using a task that did not involve an explicit spatial component \citep{Xie_2020}. Imagery and spatial attention therefore seem to share common features; they both appear to rely on top-down processing, with one consequence that perception seems to have higher spatial resolution than both spatial attention \citep{Intriligator_2001} and imagery \citep{Breedlove_2020}. It is very difficult to untangle the contributions of perceptual processes and spatial attention to internal representations. It seems likely that imagery involves mechanisms related to perception and attention, relying on top-down processing from high-level brain regions. 

One aspect that is likely to affect the top-down generation of an internal representation is how it is prompted and the ability to predict its features in advance, for example when objects become occluded. The processes underlying the representation of occluded objects may be closely related to those in conventional imagery tasks \citep{Nanay_2009}. However, there are some important differences between imagery and occlusion. Imagery can be prompted from short- or long-term memory, which involve different brain regions \citep{Ishai_2002}. Mental imagery can be considered to encompass situations in which there is a visual percept that is not produced via current sensation. In this view, representations held in working memory can therefore involve mental imagery; indeed, percepts in working memory resemble those arising from mental rotation \citep{Albers_2013}. In conditions of occlusion, as well as in the case of visual working memory, there is usually some sensory support, such as from a fragment of the object not occluded or full view of the object immediately before occlusion. One possibility is that internally generated representations utilise the same brain networks as perceptual representations, but the temporal dynamics vary with the ability to predict and anticipate details of the stimulus to be generated.

Tracking the position of a predictably moving object is a common task that may share some top-down processes with static imagery tasks. In particular, prediction is likely to play an important role in both imagery and visual tracking. The ability to predict the movement of a stimulus influences perceptual processing during visual tracking \citep{Blom_2020, Hogendoorn_2018}. Hogendoorn and Burkitt (\citeyear{Hogendoorn_2018}) measured EEG from participants who viewed an apparent motion stimulus that was predictable or unpredictable in its motion trajectory. The position-specific representations occurring 80-90ms after stimulus onset were unaffected by motion predictability, but a later stage of processing (typically 140-150ms after a stimulus is presented) occurred earlier for predictable relative to random sequences by approximately 16ms \citep{Hogendoorn_2018}. Predictability therefore has a marked effect on the temporal dynamics of spatial representations for visible stimuli. For an object appearing in an unpredictable location, the resulting position representation must be a combination of the internal representation of the expected location and the stimulus-driven response to the actual object location \citep{Blom_2020}. Disentangling a stimulus prediction from a stimulus-driven response is an important next step in understanding how and when internal representations are formed. Anticipatory mechanisms are likely to influence internally generated spatial representations, but might interact with other effects, for example the delayed processes observed during imagery \citep{Dijkstra_2018}. 

In the current study, to understand the nature of internal representations in the brain, we investigated the neural processes underlying visual tracking for visible and imagined objects. Participants covertly tracked the position of a simple moving stimulus and kept tracking its trajectory after it disappeared. Using spatial imagery allowed us to assess the temporal dynamics of internal representations during object tracking in the absence of a stimulus-driven response. EEG and time-resolved multivariate pattern analysis were used to assess the position-specific information contained within the neural signal during visible and imagined stimulus presentations. We successfully decoded the position of the stimuli from all phases of the task. Our results show that the visible and imagined stimuli evoked the same neural response patterns, but with very different temporal dynamics. Further, multivariate encoding models revealed that the spatial representations of imagined stimuli were much weaker than those of visual stimuli. These findings suggest that overlapping mid- and high-level visual processes underlie perceptual and internally generated representations of spatial location, and that these are pre-activated in anticipation of a stimulus.

\section{Methods}
All stimuli, data and analysis code are available at \url{https://osf.io/8v47t/}. The experiment consisted of two types of sequences: a pattern estimator and the experimental task. In the pattern estimator sequences, the order of the stimuli was unpredictable, whereas in the experimental task the order was predictable. The pattern estimator sequences were used to obtain position-specific EEG signals that were unlikely to be affected by eye-movements, and were subsequently used to detect position signals in the experimental task. 

\subsection{Participants}
Participants were 20 adults recruited from the University of Sydney (12 females; age range 18-52 years) in return for payment or course credit. The study was approved by the University of Sydney ethics committee and informed consent was obtained from all participants. Four participants were excluded from analyses due to excessive eye movements during the pattern estimator sequences. 

\subsection{Stimuli and design}
While participants maintained fixation in the centre of the monitor, a stimulus appeared in six distinct positions 4 degrees of visual angle from fixation. The stimulus positions were 0\degree, 60\degree, 120\degree, 180\degree, 240\degree and 300\degree relative to fixation. The stimulus was a black circle with a diameter of 3 degrees of visual angle. Six unfilled circles acted as placeholders, marking all possible positions throughout the trial. Every stimulus presentation was accompanied by a 1000Hz pure tone presented for 100ms via headphones. All stimuli were presented using Psychtoolbox \citep{Brainard_1997, Kleiner_2007, Pelli_1997} in MATLAB. In total, there were 8 blocks of trials, each of which contained two pattern estimator sequences and 36 experimental task sequences.

\subsubsection{Pattern estimator}
The pattern estimator sequences were designed to extract stimulus-driven position-specific neural patterns from the EEG signal. Participants viewed 16 pattern estimator sequences (2 per block), each of which consisted of 10 repetitions of the 6 stimulus positions (Figure \ref{Figure_1}A). The order of stimuli was randomised to ensure that for a given stimulus position, the preceding and following stimuli would not be predictive of that position; for example, comparing the neural patterns evoked by positions 1 and 2 could not be contaminated by preceding and following stimuli because they could both be preceded and followed by all six positions.  Each stimulus was shown for 100ms and was followed by an inter-stimulus interval of 200ms. Onset of the stimulus was accompanied by a 100ms tone. Participants were instructed to passively view the stimuli without moving their eyes from the fixation cross in the centre of the screen.

The stimuli were presented in unpredictable patterns so there was no regularity in the positions of the previous or following stimuli to contribute to the neural patterns extracted for each position. Additionally, the random sequences ensured that any eye movements would be irregular and thus unlikely to contribute to the extracted neural signal. Previous work has shown that even the fastest saccades typically take at least 100ms to initiate \citep{Fischer_1984}. Furthermore, eye movements do not appear to affect decoding of magnetoencephalography data until 200ms after a lateralised stimulus is presented \citep{Quax_2019}. Our 100ms stimulus duration was therefore unlikely to generate consistent eye movements that would affect the early, retinotopic EEG signal of stimulus position.

To assess whether participants complied with the fixation instruction, we assessed the EEG signal from electrodes AF7 and AF8 (located near the left and right eye, respectively) as a proxy for electrooculogram measurements. We calculated the standard deviation of the AF7 and AF8 signals across each of the 16 sequences and then averaged the deviation for the two electrodes. If a participant’s average deviation across the 16 sequences exceeded 50$\mu$V, that individual was considered to be moving their eyes or blinking too often, resulting in poor signal. An amplitude threshold of 100$\mu$V is commonly used to designate gross artefacts in EEG signal \citep{Luck_2014}, so we adopted an arbitrary standard deviation threshold of 50$\mu$V (50\% of the typical amplitude threshold) to indicate that there were too many artefacts across the entire pattern estimator sequences. Four participants exceeded this standard deviation threshold (\emph{M} = 72.72$\mu$V, range = 63.93-82.70$\mu$V) and were excluded from all analyses. For each of the remaining 16 participants, the median deviation was well below this threshold (\emph{M} = 25.92$\mu$V, SD = 5.64$\mu$V, range = 16.06-37.62$\mu$V). Thus, the four excluded participants had far more signal artefacts (probably arising from eye movements) than the other participants. 

\subsubsection{Tracking task}
\begin{figure}[h]
\centering
\includegraphics[width=12cm]{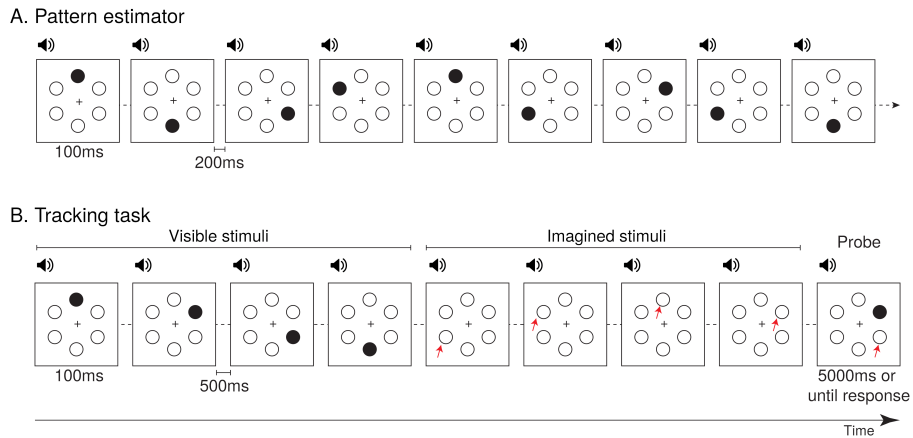}
\caption{Stimuli and design. A) Pattern estimator. Participants passively viewed rapid sequences in which a black circle stimulus appeared in six locations in random order. A tone accompanied every stimulus onset. B) Tracking task. The stimulus was presented in different locations in predictable sequences. After 4-6 visible locations, participants had to track the location of the “imagined” stimulus by imagining the continuation of the sequence. A tone accompanied every stimulus onset. During the 4-6 “imagined” positions, the auditory stimulus continued at the same rate, but only the six placeholder locations were shown. At the end of the sequence, a probe appeared, and participants had to respond if it was in the expected position or whether it was trailing or leading the sequence. This example shows a clockwise sequence with trailing probe. Red arrows (not shown in experiment) designate the expected position of the imagined stimulus.}
\label{Figure_1}
\end{figure}

For the experimental task, participants viewed sequences consisting of 4-6 visible stimuli and 4-6 “imagined” presentations simulating occluded stimuli (Figure \ref{Figure_1}). The positions of the visible stimuli were predictable, presented in clockwise or counter-clockwise sequences. Participants were asked to covertly track the position of the stimulus, and to continue imagining the sequence of positions when the stimulus was no longer visible. At the end of each sequence, there was a 1000ms blank screen followed by a probe stimulus that was presented in one of the 6 locations. Participants categorised this probe as either (1) trailing: one position behind in the sequence, (2) expected: the correct location, or (3) leading: one position ahead in the sequence. Participants responded using the Z, X or C keys on a keyboard, respectively. Each response was equally likely to be correct, so chance performance was 33.33\%.

\subsection{EEG recordings and preprocessing}
EEG data were continuously recorded from 64 electrodes arranged in the international 10–10 system for electrode placement \citep{Oostenveld_2001} using a BrainVision ActiChamp system, digitized at a 1000Hz sample rate. Scalp electrodes were referenced to Cz during recording. EEGLAB \citep{Delorme_2004} was used to pre-process the data offline, where data were re-referenced to the average of all electrodes. We filtered the data using a Hamming windowed sinc FIR filter with highpass of 0.1Hz and lowpass of 100Hz and then downsampled to 250Hz as in our previous work \citep{Grootswagers_2019, Robinson_2019}. Epochs were created for each stimulus presentation ranging from -200 to 1000ms relative to stimulus onset. No further preprocessing steps were applied. Mean neural responses of these epochs show clear event-related potentials in response to the visual and auditory stimuli (see Supplementary Material Figure \ref{S5}).

\subsection{Decoding analyses}
An MVPA decoding pipeline \citep{Grootswagers_2017} was applied to the EEG epochs to investigate position representations of visible and imagined stimuli. All steps in the decoding analysis were implemented in CoSMoMVPA \citep{Oosterhof_2016} . A leave-one-block-out (i.e., 8-fold) cross-validation procedure was used for all time-resolved decoding analyses. For each time point, a linear discriminant analysis (LDA) classifier was trained using the pattern estimator data to distinguish between all pairs of positions. LDA covariance was regularised by .01. Channel voltages from the 64 EEG channels were used as features for classification. Each classifier was trained with balanced numbers of trials per stimulus position from the pattern estimator sequences. The classifier was then tested separately on the visible and imagined positions in the experimental task. This provided decoding accuracy over time for each condition. At each time point, mean pairwise accuracy was tested against chance (50\%). Importantly, because all analyses used the randomly-ordered pattern estimator data for training the classifier, above chance classification was very unlikely to arise from the predictable sequences or eye movements in the experimental task. For the tracking task, all sequences were included in the decoding analyses regardless of whether the participant correctly classified the position of the probe (i.e., correct and incorrect sequences were analysed). When only correct trials were included, the trends in the results remained the same (see Supplementary Material  \ref{S1}).  
To assess whether neighbouring stimulus positions evoked more similar neural responses, we also calculated decoding accuracy as a function of the distance between position pairs. Each position pair had a radial distance of 60\degree, 120\degree or 180\degree apart. There were six pairs with a distance of 60\degree (e.g., 0\degree  vs 60\degree , 60\degree  vs 120\degree, 0\degree  vs 300\degree), six pairs with a distance of 120\degree (e.g., 0\degree  vs 120\degree, 60\degree  vs 180\degree), and three pairs with a distance of 180\degree (directly opposing each other, e.g., 0\degree  vs 180\degree, 60\degree  vs 240\degree). Decoding accuracy for each pair distance was calculated as the mean of all relevant pair decoding and compared to chance (50\%).

As a final set of decoding analyses, time generalisation \citep{King_2014} was used to assess whether the patterns of informative neural activity occurred at the same times for the pattern localiser and the visible and imagined stimuli on the tracking task. Classification was performed on all combinations of time points from the pattern estimator epochs and the visible or imagined epochs. Classifiers were trained on all trials from the pattern estimator sequences and tested on visible and imagined stimulus positions. 

\subsection{Multivariate encoding analyses}
As exploratory analyses prompted by reviewers’ comments, we used forward encoding models to investigate the spatial selectivity of visible and imagined representations across time. Encoding models can be used with neuroimaging data to investigate neural encoding of many visual feature dimensions \citep{Sprague_2015}. Such models have been applied to fMRI data to assess encoding of features such as colour \citep{Brouwer_2009}, orientation \citep{Scolari_2012} and position \citep{Sprague_2013}. These methods have also been adapted for use with EEG using neural responses in the frequency \citep{Foster_2016, Garcia_2013} and temporal domains \citep{Smout_2019, Tang_2018, Tang_2020}. Here, we used data from the pattern estimator task and multivariate linear regression to model the EEG responses per time point as a weighted sum of six position “channels”, each tuned to the experimental positions of 60\degree, 120\degree, 180\degree, 240\degree, 300\degree, 360\degree. These models were then used to estimate channel responses for visible and imagined positions on the tracking task, in order to assess the selectivity of the position representations. Analyses were adapted from encoding analyses of EEG data implemented in Smout, Tang, Garrido and Mattingley \citeyear{Smout_2019} using scripts on the Open Science Framework (\url{https://doi.org/10.17605/osf.io/a3pfq}) and functions from \url{https://github.com/Pim-Mostert/decoding-toolbox}. 

Results from encoding analyses are activations (rather than predictions as in decoding), so encoding is more sensitive to noise and artefacts in the data. Additional data cleaning steps were applied to remove noise and artefacts. After epoching, we interpolated electrodes that exceeded 5 standard deviations from the mean kurtosis value. For one dataset, we interpolated one additional channel that remained extremely noisy by visual inspection. In total, six or fewer channels were interpolated per dataset (<10\%, \emph{M} = 3.5, SE = .56). To remove artefacts, any epochs that exceeded +/-100$\mu$V at any time across the epoch were excluded from the analyses, and for every training/testing fold we randomly subsampled the remaining clean trials so there were equal numbers per position for the pattern estimator (total \emph{M} = 640.13, SE = 46.77) and equal numbers per position, condition (visible/imagined) and movement direction (clockwise/counter-clockwise) on the tracking task (total \emph{M} = 1957.50, SE = 177.11). These steps ensured that the position encoding analyses were based on clean EEG data and could not be biased due to unequal trial numbers.

For each participant and time point, encoding models were trained using four-fold cross-validation, each time training on 75\% of the pattern estimator data and testing on 25\% of the test data. This procedure was repeated 100 times with different trial subsampling every time \citep{Smout_2019}. These analyses resulted in response profiles across the six stimulus positions (encoding “channels”; 0, 60, 120, 180, 240 and 300\degree) for each trial. Channel responses were then realigned to positions -120\degree to 180\degree, where the 0\degree position channel reflected the correct stimulus position for the trial. We expected that the position representations on the tracking task might also include representations for the previous and next stimuli in the sequence, so we collated the data separately for clockwise and counterclockwise sequences and relabelled the position channels to reflect position relative to stimulus movement. Thus, channels +60, +120 and +180 degrees reflect positions of the next three stimuli in the sequence, and channels -60 and -120 reflect positions of the preceding two stimuli. Mean position channel responses were then calculated per time point for the visible and imagined stimuli.

To assess the position representations in the neural signal, exponentiated cosines were fit to the encoding response profiles across the six position channels for each participant, condition and time point using the equation:

\begin{equation*}
    y(x) = A * e^{\kappa(cos(x-\mu)-1)} + B
\end{equation*}

which models the expected response profile for position angle \emph{x} with a distribution with amplitude \emph{A} (peak response amplitude) with $\kappa$ concentration (sharpness of the distribution, analogous to standard deviation) that clusters around $\mu$ (peak of the function) with baseline offset \emph{B}. The fitting was implemented using lsqcurvefit in MATLAB with starting values \emph{A} = 0.2 (range -5 to 10), $\kappa$ = 1 (0 to 10), $\mu$ = 0 (-60\degree to 60\degree) and \emph{B} = 0 (-5 to 2). We analysed the amplitude \emph{A} and peak $\mu$ over time for position representations of visible and imagined stimuli.

\subsection{Statistical Inference}
To assess the evidence that decoding performance or parameter values differed from chance, we calculated Bayes factors \citep{Dienes_2011, jeffreys_1998, Kass_1995, Rouder_2009, Wagenmakers_2007}. A JZS prior \citep{Rouder_2009} was used with a scale factor of 0.707, meaning that for the alternative hypothesis of above-chance decoding, we expected to see 50\% of parameter values falling within -.707 and .707 standard deviations from chance \citep{jeffreys_1998, Rouder_2009, Wetzels_2012, Zellner_1980}. The Bayes factor (BF) indicates the probability of obtaining the group data given the alternative hypothesis relative to the probability of the data assuming the null hypothesis is true. We used thresholds of BF > 3 and BF > 10 as increasing evidence for the alternative hypothesis, and BF < 1/3 as evidence in favour of the null hypothesis \citep{jeffreys_1998, Kass_1995, Wetzels_2011}. BFs that lie between those values indicate insufficient evidence to favour either of the two hypotheses.

\section{Results}
\subsection{Behavioural results}
\begin{figure}[h]
\centering
\includegraphics[width=10cm]{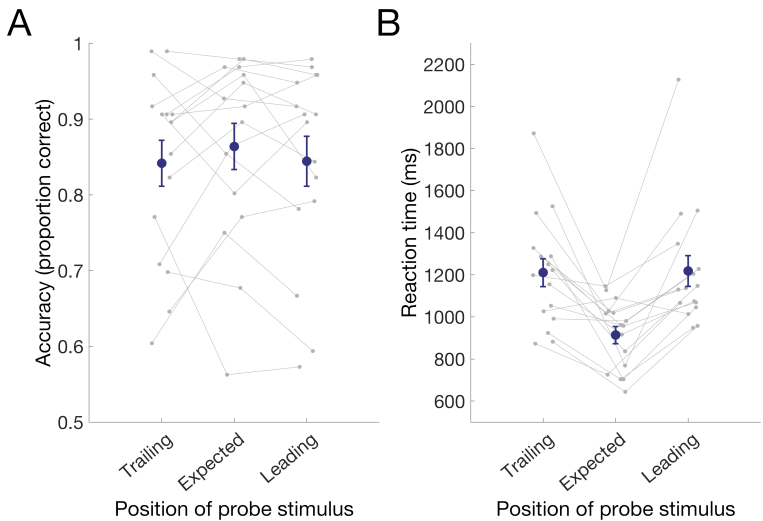}
\caption{Behavioural results. A) Accuracy, and B) Response time on the tracking task as a function of final probe position. Individual participant data are plotted in grey, with group mean in navy. Error bars depict one standard error of the mean across participants (N = 16).}
\label{Figure_2}
\end{figure}

Participants performed well on the tracking task, with high mean accuracy for all probe positions (Figure \ref{Figure_2}A). Response time was calculated within participant as the mean correct response time per probe position. At the group level, response time was faster for the expected probe position relative to the unexpected probe positions (trailing or leading) (Figure \ref{Figure_2}B). These results indicate that on most trials participants knew where the probe was meant to appear, which required tracking the expected location of the object. Evidently, participants allocated their attention appropriately to the expected position of the stimulus during the imagined portion of the tracking task.

\subsection{Position decoding using the pattern estimator sequences}
\begin{figure}[h]
\centering
\includegraphics[width=14cm]{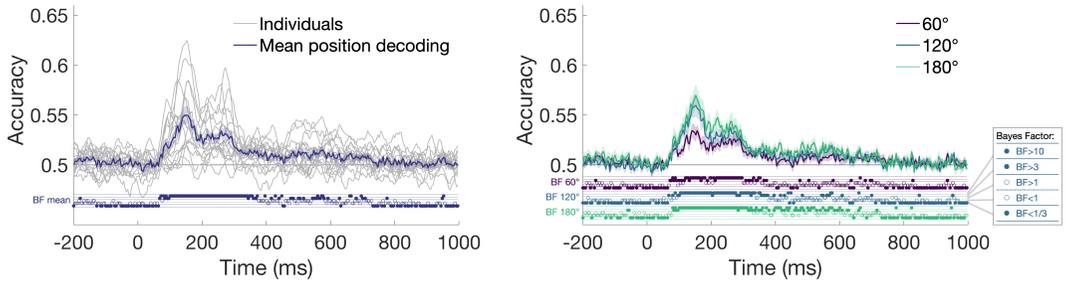}
\caption{Position decoding using pattern estimator sequences. Left plot shows group mean decoding and smoothed individual participant decoding for all pairs of positions, and right plot shows mean position decoding as a function of the angular distance between stimulus pairs. Shaded areas show standard error across participants (N = 16). Thresholded Bayes factors (BF) for above-chance decoding are displayed above the x-axes for every time point as an open or closed circle in one of four locations (see inset).}
\label{Figure_3}
\end{figure}

The pattern estimator sequences were designed to extract position-specific neural patterns of activity from unpredictable visible stimuli. Time-resolved multivariate pattern analysis (MVPA) was applied to the EEG data from the pattern estimator, which revealed that stimulus position could be decoded above chance from approximately 68ms after stimulus onset and peaked at 150ms (Figure \ref{Figure_3}), consistent with initial retinotopic processing of position in early visual areas \citep{Di_Russo_2003, Hagler_2009}. To assess how the physical distance between stimulus positions influenced the neural patterns of activity, we compared the pairwise decodability of position according to the relative angle between stimulus position pairs (i.e., angle of 60\degree, 120\degree or 180\degree between two stimulus positions). The greatest decoding performance was observed for larger angles between stimulus positions.

\subsection{Position decoding on the tracking task}
\begin{figure}[]
\centering
\includegraphics[width=14cm]{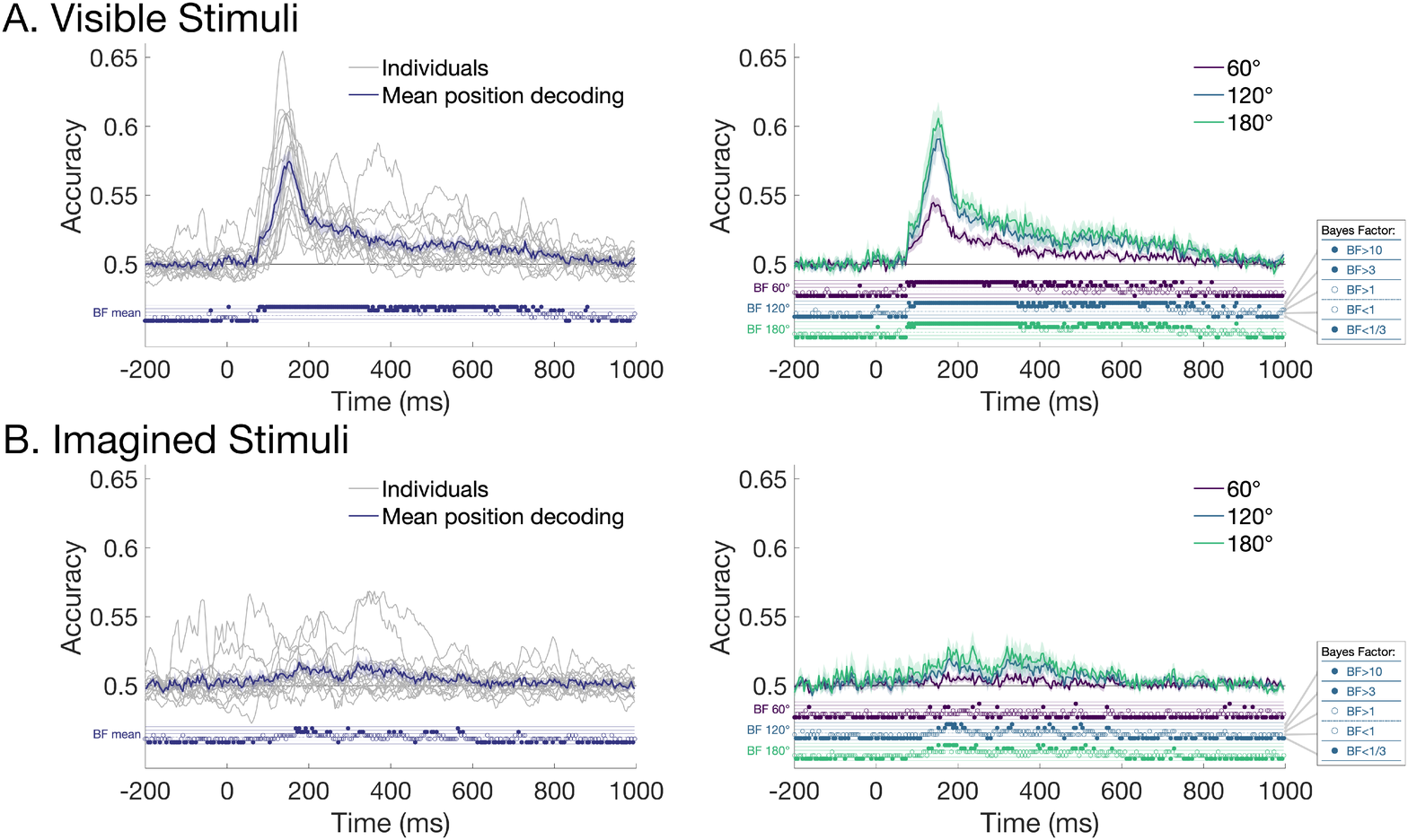}
\caption{Position decoding from object tracking task. A) Visible stimuli. B) Imagined stimuli. Left plots show group mean decoding and smoothed individual participant decoding for all pairs of positions, and right plots show mean position decoding as a function of the angular distance between position pairs. Shaded areas show standard error across participants (N = 16). Thresholded Bayes factors (BF) for above-chance decoding are displayed above the x-axes for every time point as an open or closed circle in one of four locations (see inset).}
\label{Figure_4}
\end{figure}

To assess the similarity in position representations for visible and imagined (simulated occluded) stimuli, the classifier was trained on data from the visible pattern estimator stimuli and tested on data from the tracking task for the visible and imagined stimuli. Crucially, position could be decoded for both visible and imagined stimuli, suggesting that similar neural processes underpin perceptual and internal representations of stimulus position. For visible stimuli, the pattern of decoding results echoed those of the pattern estimator, with decoding evident from approximately 76ms and peaking at 152ms, presumably reflecting visual coding of position in ventral visual areas of the brain (Figure \ref{Figure_4}A, left). When decoding was split according to the distance between the pair of positions, results looked similar to the pattern estimator results (Figure \ref{Figure_4}A, right). 

A different pattern of results was observed for the imagined stimuli. Here, decoding was not above chance until approximately 152ms and consisted of a low, broad “peak” (Figure \ref{Figure_4}B). There was considerable variation in decoding accuracy across participants (Figure \ref{Figure_4}B, left; see also Supplementary Material \ref{S6}). Although decoding accuracy was low, there was considerable evidence that accuracy was above chance (see Supplementary Material \ref{Figure_S2} for Bayes Factors in more detail). Reliable above chance cross-decoding from the visible pattern estimator stimuli to the imagined stimuli on the tracking task indicates that overlapping processes underlie stimulus-driven and internally-generated representations of spatial location. But this decoding of the internal representation of position was later and less accurate than position decoding for visible stimuli. Similar to the pattern estimator and visible decoding results, positions that were further apart were more decodable (Figure \ref{Figure_4}B, right). Notably, neighbouring positions (60\degree apart) showed little evidence of position decoding, suggesting that the representations of position were spatially diffuse for the imagined stimuli, unlike for the visible stimuli. 

\begin{figure}[]
\centering
\includegraphics[width=12cm]{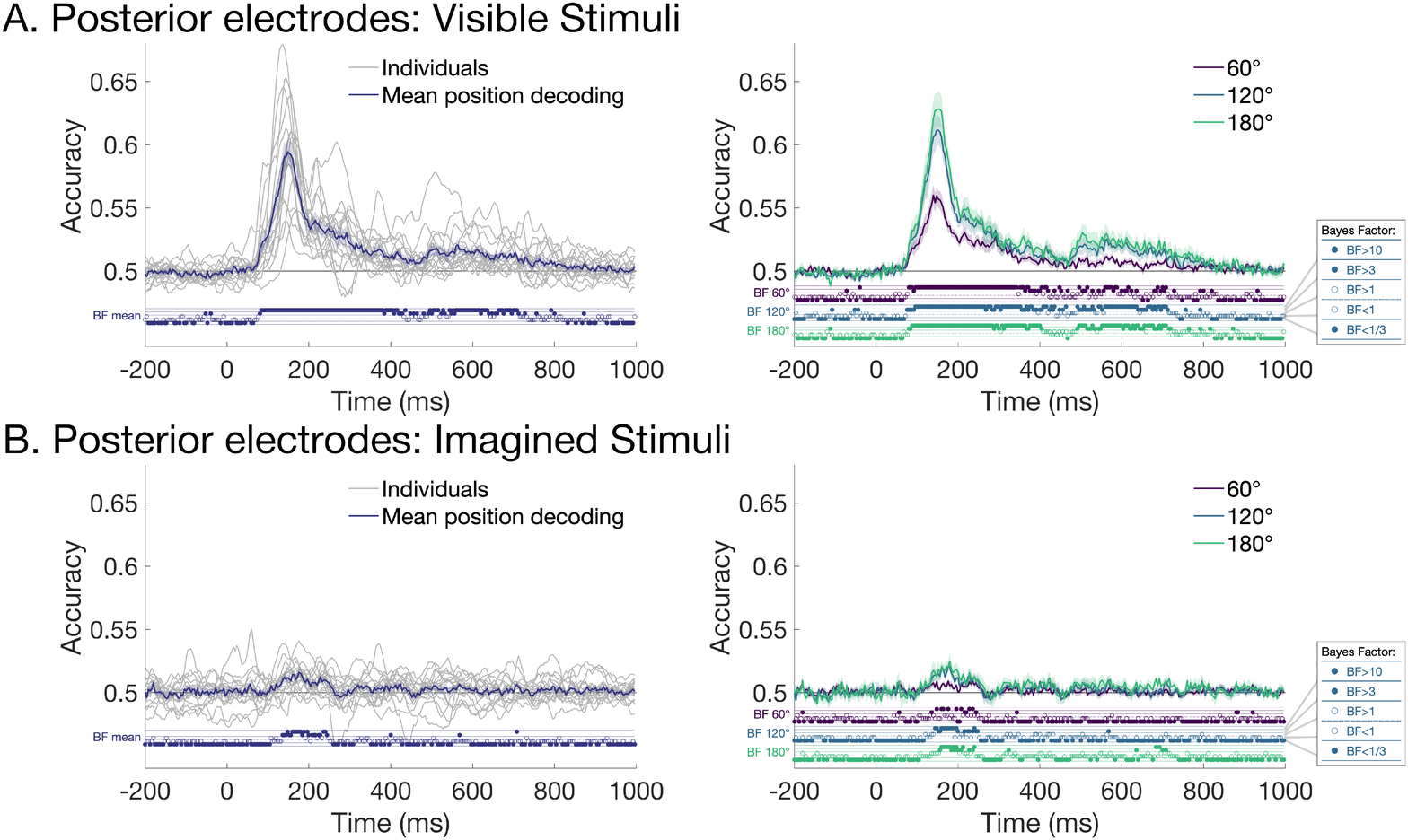}
\caption{Position decoding from object tracking task using only posterior electrodes. A) Visible stimuli. B) Imagined stimuli. Left plots show group mean decoding and smoothed individual participant decoding for all pairs of positions, and right plots show mean position decoding as a function of the angular distance between stimulus pairs. Shaded areas show standard error across participants (N = 16). Thresholded Bayes factors (BF) for above-chance decoding are displayed above the x-axes for every time point as an open or closed circle in one of four locations (see inset).}
\label{Figure_5}
\end{figure}

The previous analyses were performed using electrodes covering the whole head, which meant that there was a possibility that non-neural artefacts such as eye movements might contribute to the classification results \citep{Quax_2019}. Saccadic artefacts tend to be localised to frontal electrodes, close to the eyes \citep{Lins_1993}. To assess if the EEG signal contributing to the position-specific neural information originated from posterior regions of the brain (e.g., occipital cortex), as expected, we conducted the same time-resolved decoding analyses using a subset of electrodes from the back half of the head. We used 28 electrodes that were likely to pick up the largest signal from occipital, temporal and parietal areas (and were less likely to be contaminated with frontal or muscular activity). The electrodes were CPz, CP1, CP2, CP3, CP4, CP5, CP6, Pz, P1, P2, P3, P4, P5, P6, P7, P8, POz, PO3, PO4, PO7, PO8, Oz, O1, O2, TP7, TP8, TP9 and TP10. As can be seen in Figure \ref{Figure_5}, the same trend of results was seen using this subset of electrodes compared with the whole head analyses in Figure \ref{Figure_4}. Specifically, Bayes Factors revealed evidence that position of imagined stimuli was decodable approximately 136-244ms, which is slightly earlier than the whole brain results. Decoding was also most evident for positions that were a distance of 120\degree or 180\degree apart (Figure \ref{Figure_5}B). Interestingly, imagery decoding was more prolonged for the whole-brain decoding than posterior analyses, which could reflect higher-order cognitive processing of stimulus position in more anterior regions of the brain, or increased power due to more features (electrodes) included in the whole brain analysis. Analyses restricted to frontal electrodes showed later, more diffuse coding for visible stimuli relative to the posterior analysis, and little evidence for position coding of imagined stimuli (see Supplementary Material \ref{S3}). Thus, position-specific neural information for visible and imagined stimuli was evident specifically over posterior regions of the brain, consistent with visual cortex representing stimulus-driven and internal representations of spatial location.

\begin{figure}[h]
\centering
\includegraphics[width=14cm]{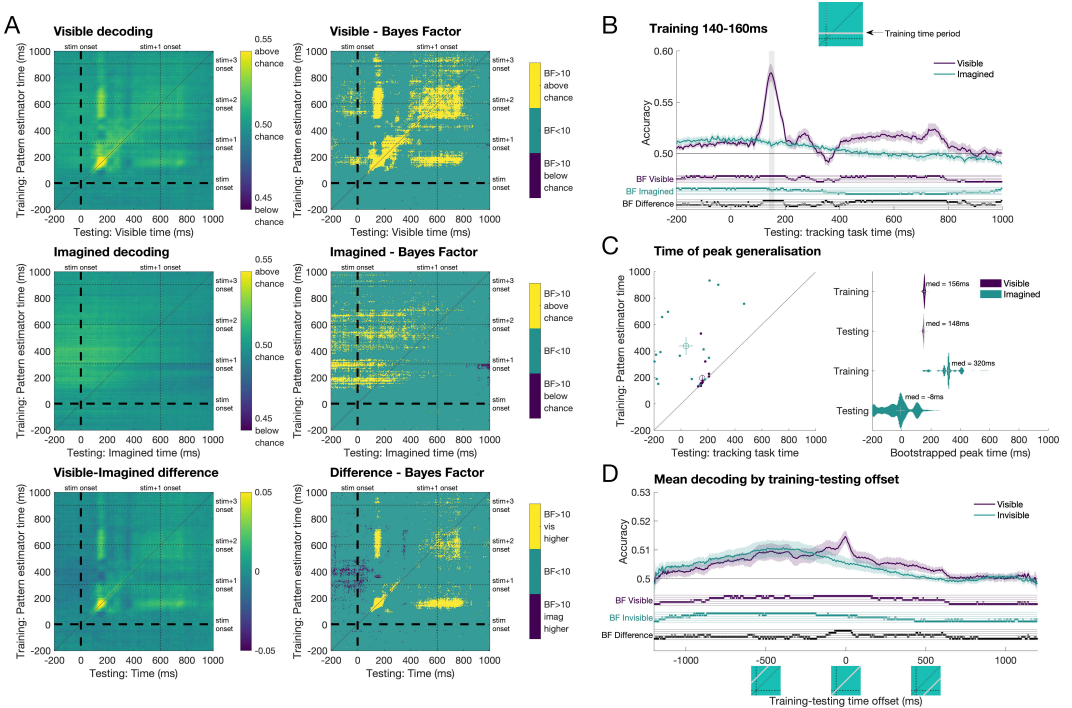}
\caption{Time generalisation results. A) Decoding stimulus position for visible stimuli and imagined stimuli. Left plots show decoding for visible, imagined and visible-imagined difference, and right plots show associated Bayes Factors. Decoding was performed by training on data from the pattern estimator sequences of visible stimuli and testing on the experimental trials, for all pairs of time points. B) Decoding accuracy using training times 140-160ms on the pattern estimator and testing all time points for visible and imagined stimuli. C) Peak decoding times for training and testing processes. Left plot shows peak decoding times per participant, and right plot shows distribution of peak times after bootstrapping the group 1000 times. D) Mean decoding accuracy for different training-testing time offsets. Highest decoding for visible stimuli occurred around 0ms offset, indicating processes occurred at the same time points for the pattern estimator and visible stimuli on the tracking task. In comparison, the highest decoding for imagined stimuli occurred earlier in the test set than the training set.}
\label{Figure_6}
\end{figure}

The results of the time-resolved analyses showed that position-specific neural patterns for visible stimuli generalised to imagined stimuli, but with different temporal dynamics. To assess the possibility that neural processes were more temporally variable for imagined than for visible stimuli, we performed whole brain (64-channel) time-generalisation analyses by training the classifier on all time points of the pattern estimator and testing on all time points from the tracking task. As expected, position could be decoded from both visible and imagined stimulus presentations, but with marked differences in their dynamics (Figure \ref{Figure_6}). For the visible stimuli, most of the above-chance decoding was symmetric on the diagonal, indicating that the position-specific processes occurred at approximately the same time for visible stimuli in the pattern localiser and the tracking task (Figure \ref{Figure_6}A, top), even though the inter-stimulus intervals for stimuli in the training and test sets were different. Interestingly, there was also some above-diagonal decoding indicating that some neural signals observed in the pattern localiser occurred substantially earlier in the tracking task, which may reflect prediction based on the previous stimuli. Also likely reflecting anticipation of the stimulus position, generalisation occurred for time points prior to onset of the visible stimulus in the tracking task. About 800-1000ms after the tracking stimulus was presented, there is some evidence of below chance decoding, indicating a different stimulus position was systematically predicted. This is likely to reflect processing of the next stimulus in the tracking task, which was presented at 600ms on the plot (stim +1 vertical line).  

Time generalisation for the imagined stimulus position was not centred on the diagonal, reflecting different temporal dynamics for the predicted internal representations than for the stimulus-driven processing of the pattern estimator. Decoding generalisation was also much more diffuse and relied on processes approximately 120-750ms after stimulus onset in the pattern estimator (Figure \ref{Figure_6}A, middle). Decoding again preceded the onset of the tone in the tracking task, reflecting an anticipation effect. There was also below chance decoding at later time points, indicating that the classifier was predicting a different stimulus position at times when the next stimulus would be processed. Comparison between visible and imagined position showed higher decoding for the imagined stimuli preceding the tone, but higher decoding for the visible stimuli after the stimulus and tone were presented (Figure \ref{Figure_6}A, bottom). 

The dynamics of the time generalisation results give insight into the processing underlying perceptual and imagined position representations. Using decoding models trained on the pattern estimator at 140-160ms (approximately the time of peak position decoding), we looked at decoding accuracy for each time point on the tracking task. It is clear that visible representations show stimulus-evoked position specific responses, with largest decoding at the same time period as the training times (Figure \ref{Figure_6}B). Imagined representations, however, show much more diffuse responses that ramp up earlier than those of visible stimuli, with imagined decoding highest before 0ms, the time of the tone. Interestingly, this plot resembled within-condition decoding results (i.e., training and testing on visible or imagined stimuli from the training task; see Supplementary Material \ref{S4}). 

The time generalisation results show that position representations seem to emerge earlier for imagined than visible stimuli. For peak decoding times per participant (Figure \ref{Figure_6}C, left), visible position was most separable when training and testing approximately the same time points (about 150ms), whereas imagined position relied on later training than testing times, and showed much more variability across participants. To further assess peak decoding times, we bootstrapped the group 1000 times with replacement and calculated the times of peak generalisation to assess the distribution. Figure \ref{Figure_6}C (right) shows that visible decoding showed training and testing peaks at approximately 150ms with very little variation across the 1000 iterations. Imagined representations, by contrast, peaked after 300ms for training and 0ms for testing. Finally, assessing decoding accuracy by training-testing lag revealed that imagined decoding was higher when training on later time points than testing times, whereas visible decoding was highest at approximately 0ms offset (i.e., same training and testing times; Figure \ref{Figure_6}D). These results suggest that imagined representations rely on high level perceptual and cognitive processes that are implemented earlier in time. Overall, the time generalisation results suggest that during the imagined stimulus portion of the tracking task, which relied on internal representations of position, the neural dynamics were more anticipatory and variable than perceptual processes.

\begin{figure}[h]
\centering
\includegraphics[width=14cm]{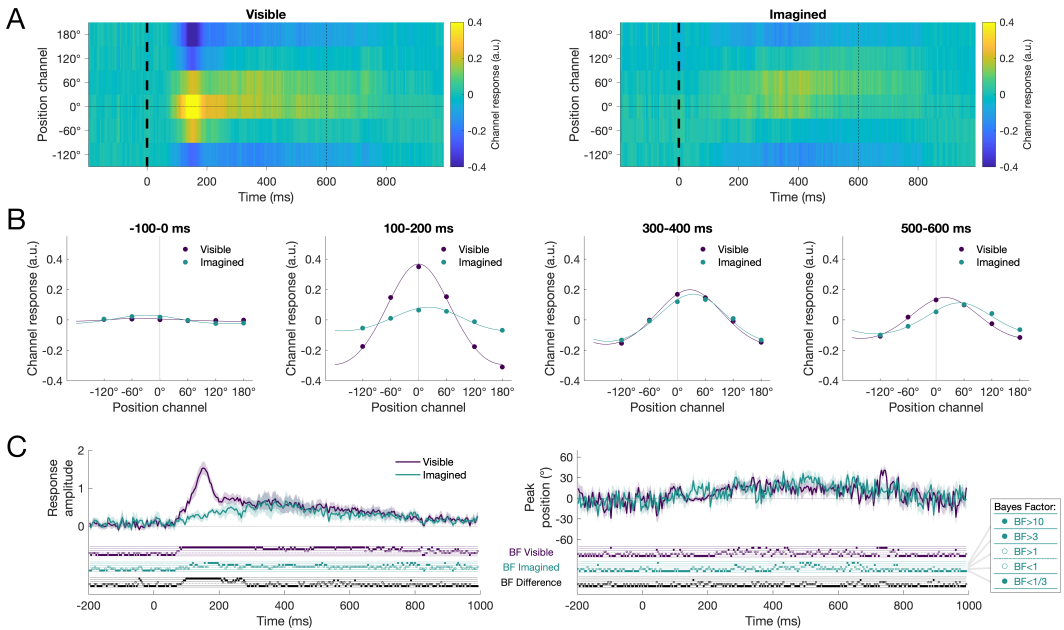}
\caption{Position response profiles for visible and imagined stimuli using encoding models trained on the random pattern estimator stimuli. A) Activations of each encoding position channel for visible and imagined stimuli, plotted as relationship to the presented stimulus position. For the visible and imagined conditions, there was higher activation for position channels closer to the correct position (0\degree), indicating that the neural representation of stimulus position was captured by the encoding model. B) Model fitting of channel responses for some representative time periods show the emergence of spatial information over time, with a shift towards the next stimulus position. C) Parameters of the model fits over time. Left: Response amplitude at each time point for visible and imagined stimuli. There was reliable spatial signal for both conditions, but the temporal dynamics varied. Right: Peak position for the model fit generally reflected a peak at 0 degrees for visible and imagined stimuli, although after 250ms there was some evidence of positive shifts towards the upcoming stimulus position.}
\label{Figure_7}
\end{figure}

\subsection{Encoding analyses}
We used forward encoding to assess the spatial representations in the neural signal for visible and imagined stimuli. Encoding models were trained on EEG data from the pattern estimator sequences and applied to the tracking task separately for each condition and time point. This exploratory analysis resulted in activations per condition for six encoding channels representing the experimental positions. Figure \ref{Figure_7}A depicts mean response profiles per condition, which show higher responses for the actual stimulus position (0\degree) relative to the other positions, indicating that visible and imagined position were encoded according to the same processes as the pattern estimator. Plots of the response profiles at representative time periods show position information emerges over time but appears to shift towards to next stimulus position in the sequence (denoted by 60\degree) at later time periods (Figure \ref{Figure_7}B). Fitting a model to the channel responses at each time point resulted in two relevant parameters of the spatial coding of the neural signal. The amplitude of the model fit, the peak response amplitude, emerged over time and was reliably above zero for both visible and imagined stimuli, although it emerged slightly later for imagined stimuli (Figure \ref{Figure_7}C, left). This plot resembled position decoding over time. Importantly, amplitude was reliably higher for visible than imagined position from approximately 100-276ms, indicating that the neural representations of position are stronger for physical rather than internally generated stimuli. Modelling of the response profiles also revealed for peak position (i.e., the centre of the model fit) there was some evidence of a shift away from the current stimulus position in the positive direction for both visible and imagined stimuli (Figure \ref{Figure_7}C, right). This positive shift was evident from about 300ms, suggesting that the position representations at this time were more consistent with the upcoming stimulus position. Together, these encoding analyses complement the decoding results by showing that visible and imagined spatial position are encoded using stimulus-driven processes, and that imagined stimuli elicit considerably weaker spatial representations than visible stimuli.

\section{Discussion}
In this study, we assessed the neural underpinnings of internally-generated representations of spatial location. Participants viewed predictable sequences of a moving stimulus and imagined the sequence continuing when the stimulus disappeared. Time-resolved multivariate analyses revealed that patterns of activity associated with visual processing in random sequences were also associated with processing of visible and imagined spatial stimulus positions in the tracking task, but with different temporal dynamics. Specifically, the neural correlates of imagined position (i.e., internally-generated representations) were anticipatory and more temporally diffuse than those of visible position (i.e., sensory-driven representations). Taken together, this study provides evidence that internal representations of spatial position rely on mechanisms of visual processing, but that these are applied with different temporal dynamics to actual perceptual processes.

The results of this study suggest that similar perceptual and cognitive processes are implemented for processing position of visible and imagined (e.g., occluded) stimuli. This adds to previous neuroimaging work using high level objects by showing that internally-generated spatial representations appear to use the same visual perceptual processes as viewed stimuli \citep{Dijkstra_2018}. What neural processes are responsible for this low-level spatial imagery? We found generalisation from the pattern estimator to the visible tracked stimuli began at approximately 76ms, but for imagined stimuli the generalisation did not occur until 120ms (Figure \ref{Figure_6}A). This suggests that internal spatial representations do not originate with early retinotopic processes such as that of the first stages of processing in V1, but are implemented by higher order processes potentially via feedback and recurrent processes. Above-chance generalisation for visible and imagined stimuli was maintained until approximately 750ms after the pattern estimator stimulus was presented, indicating that position-specific information represented throughout the visual hierarchy has some similarity for stimulus-driven and internally generated representations. It is important to note, however, that the time generalisation results did not show evidence of distinct, progressive stages of processing for the imagined representations. In contrast, the visible stimuli showed different clusters of above-chance decoding on the diagonal of the time-generalisation results, indicating that there were distinct stages of processing.  These results are similar to those observed in Dijkstra et al., \citeyear{Dijkstra_2018} during imagery of faces and houses. Recent work has suggested that imagery involves a flow of information from higher- to lower-level brain regions in succession through the ventral stream \citep{Breedlove_2020}. We did not find any evidence of this reversal of perceptual processes in imagery; rather, our results suggest that internal representations activate different perceptual stages simultaneously. However, new analysis methods might yield more insight into the information flow through different brain regions during imagery \citep{Dijkstra_2019}.

For both visible and imagined stimuli, more distant stimulus positions could more easily be discriminated by the EEG signals. Decoding for neighbouring positions (60\degree apart) was generally much lower than decoding for positions that were further apart. This is consistent with the retinotopic organization of visual cortices \citep{Tootell_1998}, where closer areas of space are represented in neighbouring regions of cortex, leading to more similar spatial patterns of activation that are measured on the scalp with EEG \citep{Carlson_2011}. Time generalisation results also showed that neural patterns of activity from the pattern estimator sequences generalised to neighbouring positions in the visible condition, highlighting the neural similarity for close spatial representations (see $\sim$750ms in Figure \ref{Figure_6}B). Interestingly, however, decoding for the closest positions was particularly low for the imagined stimuli, indicating that internally generated representations of position are more spatially diffuse than perceptual representations. Multivariate encoding analyses verified that the neural representations of spatial position for both visible and imagined stimuli were encoded using stimulus-driven processes (modelled using data from the pattern estimator sequences), but that imagined representations of position were weaker than stimulus-driven representations from 100-276ms. Weaker spatial signal in imagery is consistent with representations originating in higher-level regions of the visual hierarchy, which have larger receptive fields \citep{Breedlove_2020}. Together, these results suggest that there are common, retinotopic mechanisms for processing position of both visible and imagined stimuli, but with important differences in the origin of the representations leading to much greater precision for visible stimuli.

A cognitive process that might contribute to the extracted position-specific signal in the current study is that of spatial attention. In our experimental task, participants were explicitly asked to track the position of the stimulus, and they performed well, suggesting they were directing their attention to the location of the stimulus. Spatial attention influences the amplitude of early EEG responses (for review, see \citet{MANGUN_1995}), and MEG classification work has shown that spatial attention enhances object decoding at early stages of processing \citep{Goddard_2019}. Top-down spatial attention also results in more diffuse spatial representations than stimulus processing \citep{Intriligator_2001}. Our decoding and encoding results were obtained from training on the pattern estimator, so our results are focused on processes common to the pattern estimator and the tracking task. In the pattern estimator, there was no explicit task and therefore no incentive to specifically attend to stimulus position. However, there was only one stimulus presented at a time and the saliency of the onsets were likely to attract attention, albeit in a different fashion to the cued positions in the experimental tracking task. As such, the pattern estimator and tracking task had different spatial attention demands, but that does not rule out spatial attention as a source of overlap between the two types of sequences. It is difficult to untangle perceptual and attentional mechanisms during imagery, and it is possible that internal spatial representations rely on processes that are common to perception and attention. The current results are consistent with previous work on imagery using paradigms that are unlikely to rely on spatial attention \citep{Dijkstra_2018, Xie_2020}, so it seems likely that perceptual mechanisms are at least a considerable source of overlap for neural patterns on the random pattern estimator sequences and the imagined positions on the tracking task. Future work could attempt to disentangle the role of perceptual and attentional processes in spatial imagery with a manipulation to reduce attention during the pattern estimator or using valid and invalid cues for spatial position.

To investigate the neural processes underlying spatial imagery, this study focused on spatial representations that were common to two different types of sequences: the pattern estimator and the tracking task. All analyses were performed by obtaining patterns of neural activity associated with spatial position from the randomly ordered pattern estimator stimuli and assessing how these patterns are similar to the position representations of visible and imagined stimuli during the tracking task. Training encoding and decoding models on an independent task allowed us to draw conclusions about the nature of the position representations during tracking without the confound of prediction. Our results show that spatial imagery implements similar neural processes as viewing stimuli. However, it is possible that spatial imagery also contains different information to the pattern estimator. As an exploratory analysis, we investigated the temporal dynamics of position representations by decoding within condition (visible and imagined) on the tracking task (see Supplementary Material \ref{S4}). This cross-validated decoding is somewhat problematic due to the predictable nature of the sequence, so decoding is above chance throughout the whole time period. However, the dynamics are still informative; specifically, within-condition decoding revealed very similar dynamics to the original decoding analysis. Within-visible decoding had a peak at 150ms, resembling the time-resolved analyses from training the pattern estimator and testing the visible stimuli on the same time points (as in Figure \ref{Figure_4}A). Within-imagined decoding was highest around 0ms, resembling the time-generalisation results from training on mid- and high-level processes of the pattern estimator (e.g., Figure \ref{Figure_6}B). These results suggest that neural processes as measured in the pattern estimator do capture most of the relevant neural processes implemented during stimulus tracking for visible and imagined stimuli. 

Spatial imagery representations were evident using multivariate decoding and encoding analyses, but the magnitudes of the effects were very small. One likely contributing factor to the small effects is the temporal jitter in the neural representations evoked by imagery both within and across participants. Time-locked analyses assess reliable patterns of neural activity occurring at the exact same time across trials. Imagery, as an internally-generated process, is likely to be much more temporally variable than perception, resulting in smaller, more diffuse time-locked neural signals. There is also the likelihood that different participants will use different strategies, resulting in variation from participant to participant. Temporal variability is a challenge in all research involving mental imagery. Indeed, in a face versus house imagery experiment, Dijkstra and colleagues (\citeyear{Dijkstra_2018}) found <60\% accuracy for imagined decoding compared with nearly 90\% for viewed stimuli. To minimise temporal variation in imagery in the current experiment, we used tones to guide participants in the timing of the task. Participants had to covertly track stimulus position on thousands of trials (>1400 visible + >1400 imagined per participant; see Supplementary Material \ref{S1} for details), and this large number of trials ensured that we had the power to capture the neural processes associated with spatial imagery representations, despite the temporal variation. Decoding accuracy was low, but accuracy is not an effect size \citep{Hebart_2018}. Our analyses show there were reliable spatial imagery representations that shared neural patterns with stimulus-driven representations. Importantly, temporal jitter for imagery cannot explain the observed temporal dynamics for processing of imagined position (as seen in the time generalisation plots in Figure \ref{Figure_6}), because jitter would predict only the x-axis of the time-generalisation plots being smeared relative to the visible condition. The observed imagery results appear to be diffuse in terms of the contributions of the pattern estimator (training; y-axis) processes, reflecting processing occurring at different times in visible and imagined parts of the task. 

One factor that we tried to control in this study was eye movements. Recent work has shown that even when participants were instructed to maintain central fixation, the spatial position of a peripheral stimulus could be decoded from eye movements, and the eye movements appeared to account for variance in the MEG signal from 200ms after the stimulus was presented \citep{Quax_2019}. To reduce the likelihood of eye movements influencing our spatial representation results, one countermeasure we implemented was using independent sequences of randomly ordered visible stimuli (pattern estimator sequences) to extract position-specific patterns from the EEG signal and used these to generalise to the tracking task. Thus, only neural signals in common between the pattern estimator and the tracking task could result in above chance decoding. The position sequences in the pattern estimator (training set) were randomised, so any incidental eye movements were unlikely to consistently vary with position. The tracking task implemented both clockwise and counter-clockwise sequences, so if there were eye movements, across the whole experiment a given position would have two completely different eye movement patterns. Above-chance cross-decoding from the pattern estimator to the tracking task was therefore unlikely to be driven by eye movements. Second, all stimuli were presented briefly (100ms duration), and for a short 200ms inter-stimulus interval during the pattern estimator. This rapid presentation rate reduced the likelihood that participants would overtly move their eyes, as even the fastest saccades take at least 100ms to initiate \citep{Fischer_1984}. Third, we excluded participants that appeared to move their eyes excessively during the pattern estimator sequences, which were the sequences used for training the classifier. Finally, we conducted an additional analysis using only posterior electrodes to validate that the neural patterns of activity informative for spatial position were consistent with processes within the visual system (e.g., from occipital cortex). Decoding from posterior electrodes was similar to the whole-brain results. Furthermore, a similar analysis using only frontal electrodes showed later, more diffuse position decoding for visible stimuli, and insufficient evidence for position decoding of imagined stimuli (see Supplementary Material \ref{S3}), indicating that frontal signal or artefacts did not drive decoding of spatial position for visible or imagined stimuli. Taken together, our finding that spatial position generalised from the pattern estimator to the tracking task from relatively early stages of processing indicates that it was actually a neural representation of spatial location that was driving the classifier rather than any overt eye movements.

In conclusion, in this study we successfully showed that the position of predictable visible and imagined stimuli can be modelled using patterns of neural activity extracted from independent visible stimuli. Our findings suggest that internally generated spatial representations involve mid- and high-level perceptual processes. The visible stimuli that we used relied on early retinotopic visual processes, yet we found no evidence of generalisation from very early processes (90-120ms) to the imagined stimuli. The stimuli we used were much simpler than the vivid, complex objects used in previous work, but we found similar stages of processing generalised from perceptual to internally-generated representations \citep{Dijkstra_2018}, suggesting a general role of mid- and high-level perceptual processing in internally-generated representations such as those implemented during imagery or occlusion. Our finding that neural representations of spatial location were weaker and occurred earlier for imagined objects than for the unpredictable objects indicates an important role of prediction in generating internal representations. Together, our findings suggest that similar neural mechanisms underlie internal representations and stimulus-driven mechanisms, but the timing of these processes is dependent on the predictability of the stimulus.

\section*{Acknowledgements}
This research was supported by an Australian Research Council Discovery Early Career Research Award (DE200101159) to A.K.R. and Australian Research Council Discovery Projects (DP160101300 and DP200101787) to T.A.C. The authors acknowledge the Sydney Informatics Hub and the University of Sydney’s high performance computing cluster Artemis for providing the high performance computing resources that contributed to these research results. We thank Alexander Sulfaro for insightful discussions about the study.

\printendnotes

\bibliography{References.bib}

\begin{thebibliography}{59}
\expandafter\ifx\csname natexlab\endcsname\relax\def\natexlab#1{#1}\fi
\expandafter\ifx\csname url\endcsname\relax
  \def\url#1{\texttt{#1}}\fi
\expandafter\ifx\csname urlprefix\endcsname\relax\def\urlprefix{URL: }\fi

\bibitem[{Albers et~al.(2013)Albers, Kok, Toni, Dijkerman and
  de~Lange}]{Albers_2013}
Albers, A.~M., Kok, P., Toni, I., Dijkerman, H.~C. and de~Lange, F.~P. (2013)
  Shared representations for working memory and mental imagery in early visual
  cortex.
\newblock \textit{Current Biology}, \textbf{23}, 1427--1431.
\newblock \urlprefix\url{https://doi.org/10.1016\%2Fj.cub.2013.05.065}.

\bibitem[{Blom et~al.(2020)Blom, Feuerriegel, Johnson, Bode and
  Hogendoorn}]{Blom_2020}
Blom, T., Feuerriegel, D., Johnson, P., Bode, S. and Hogendoorn, H. (2020)
  Predictions drive neural representations of visual events ahead of incoming
  sensory information.
\newblock \textit{Proceedings of the National Academy of Sciences},
  \textbf{117}, 7510--7515.
\newblock \urlprefix\url{https://doi.org/10.1073\%2Fpnas.1917777117}.

\bibitem[{Brainard(1997)}]{Brainard_1997}
Brainard, D.~H. (1997) The psychophysics toolbox.
\newblock \textit{Spatial Vision}, \textbf{10}, 433--436.
\newblock \urlprefix\url{https://doi.org/10.1163\%2F156856897x00357}.

\bibitem[{Breedlove et~al.(2020)Breedlove, St-Yves, Olman and
  Naselaris}]{Breedlove_2020}
Breedlove, J.~L., St-Yves, G., Olman, C.~A. and Naselaris, T. (2020) Generative
  feedback explains distinct brain activity codes for seen and mental images.
\newblock \textit{Current Biology}, \textbf{30}, 2211--2224.e6.
\newblock \urlprefix\url{https://doi.org/10.1016\%2Fj.cub.2020.04.014}.

\bibitem[{Brouwer and Heeger(2009)}]{Brouwer_2009}
Brouwer, G.~J. and Heeger, D.~J. (2009) Decoding and reconstructing color from
  responses in human visual cortex.
\newblock \textit{Journal of Neuroscience}, \textbf{29}, 13992--14003.
\newblock \urlprefix\url{https://doi.org/10.1523\%2Fjneurosci.3577-09.2009}.

\bibitem[{Carlson et~al.(2011)Carlson, Hogendoorn, Kanai, Mesik and
  Turret}]{Carlson_2011}
Carlson, T.~A., Hogendoorn, H., Kanai, R., Mesik, J. and Turret, J. (2011) High
  temporal resolution decoding of object position and category.
\newblock \textit{Journal of Vision}, \textbf{11}, 9--9.
\newblock \urlprefix\url{https://doi.org/10.1167\%2F11.10.9}.

\bibitem[{Delorme and Makeig(2004)}]{Delorme_2004}
Delorme, A. and Makeig, S. (2004) {EEGLAB}: an open source toolbox for analysis
  of single-trial {EEG} dynamics including independent component analysis.
\newblock \textit{Journal of Neuroscience Methods}, \textbf{134}, 9--21.
\newblock \urlprefix\url{https://doi.org/10.1016\%2Fj.jneumeth.2003.10.009}.

\bibitem[{Dentico et~al.(2014)Dentico, Cheung, Chang, Guokas, Boly, Tononi and
  Veen}]{Dentico_2014}
Dentico, D., Cheung, B.~L., Chang, J.-Y., Guokas, J., Boly, M., Tononi, G. and
  Veen, B.~V. (2014) Reversal of cortical information flow during visual
  imagery as compared to visual perception.
\newblock \textit{{NeuroImage}}, \textbf{100}, 237--243.
\newblock \urlprefix\url{https://doi.org/10.1016\%2Fj.neuroimage.2014.05.081}.

\bibitem[{{Di Russo}(2003)}]{Di_Russo_2003}
{Di Russo}, F. (2003) Source analysis of event-related cortical activity during
  visuo-spatial attention.
\newblock \textit{Cerebral Cortex}, \textbf{13}, 486--499.
\newblock \urlprefix\url{https://doi.org/10.1093\%2Fcercor%2F13.5.486}.

\bibitem[{Dienes(2011)}]{Dienes_2011}
Dienes, Z. (2011) Bayesian versus orthodox statistics: Which side are you on?
\newblock \textit{Perspectives on Psychological Science}, \textbf{6}, 274--290.
\newblock \urlprefix\url{https://doi.org/10.1177\%2F1745691611406920}.

\bibitem[{Dijkstra et~al.(2019)Dijkstra, Ambrogioni and van
  Gerven}]{Dijkstra_2019}
Dijkstra, N., Ambrogioni, L. and van Gerven, M.~A. (2019) Neural dynamics of
  perceptual inference and its reversal during imagery.
\newblock \urlprefix\url{https://doi.org/10.1101\%2F781294}.

\bibitem[{Dijkstra et~al.(2018)Dijkstra, Mostert, de~Lange, Bosch and van
  Gerven}]{Dijkstra_2018}
Dijkstra, N., Mostert, P., de~Lange, F.~P., Bosch, S. and van Gerven, M.~A.
  (2018) Differential temporal dynamics during visual imagery and perception.
\newblock \textit{{eLife}}, \textbf{7}.
\newblock \urlprefix\url{https://doi.org/10.7554\%2Felife.33904}.

\bibitem[{Dijkstra et~al.(2017)Dijkstra, Zeidman, Ondobaka, van Gerven and
  Friston}]{Dijkstra_2017}
Dijkstra, N., Zeidman, P., Ondobaka, S., van Gerven, M. A.~J. and Friston, K.
  (2017) Distinct top-down and bottom-up brain connectivity during visual
  perception and imagery.
\newblock \textit{Scientific Reports}, \textbf{7}.
\newblock \urlprefix\url{https://doi.org/10.1038\%2Fs41598-017-05888-8}.

\bibitem[{Fischer and Ramsperger(1984)}]{Fischer_1984}
Fischer, B. and Ramsperger, E. (1984) Human express saccades: extremely short
  reaction times of goal directed eye movements.
\newblock \textit{Experimental Brain Research}, \textbf{57}.
\newblock \urlprefix\url{https://doi.org/10.1007\%2Fbf00231145}.

\bibitem[{Foster et~al.(2016)Foster, Sutterer, Serences, Vogel and
  Awh}]{Foster_2016}
Foster, J.~J., Sutterer, D.~W., Serences, J.~T., Vogel, E.~K. and Awh, E.
  (2016) The topography of alpha-band activity tracks the content of spatial
  working memory.
\newblock \textit{Journal of Neurophysiology}, \textbf{115}, 168--177.
\newblock \urlprefix\url{https://doi.org/10.1152\%2Fjn.00860.2015}.

\bibitem[{Garcia et~al.(2013)Garcia, Srinivasan and Serences}]{Garcia_2013}
Garcia, J.~O., Srinivasan, R. and Serences, J.~T. (2013) Near-real-time
  feature-selective modulations in human cortex.
\newblock \textit{Current Biology}, \textbf{23}, 515--522.
\newblock \urlprefix\url{https://doi.org/10.1016\%2Fj.cub.2013.02.013}.

\bibitem[{Goddard et~al.(2019)Goddard, Carlson and Woolgar}]{Goddard_2019}
Goddard, E., Carlson, T.~A. and Woolgar, A. (2019) Spatial and
  feature-selective attention have distinct effects on population-level tuning.
\newblock \urlprefix\url{https://doi.org/10.1101\%2F530352}.

\bibitem[{Grootswagers et~al.(2019)Grootswagers, Robinson and
  Carlson}]{Grootswagers_2019}
Grootswagers, T., Robinson, A.~K. and Carlson, T.~A. (2019) The
  representational dynamics of visual objects in rapid serial visual processing
  streams.
\newblock \textit{{NeuroImage}}, \textbf{188}, 668--679.
\newblock \urlprefix\url{https://doi.org/10.1016\%2Fj.neuroimage.2018.12.046}.

\bibitem[{Grootswagers et~al.(2017)Grootswagers, Wardle and
  Carlson}]{Grootswagers_2017}
Grootswagers, T., Wardle, S.~G. and Carlson, T.~A. (2017) Decoding dynamic
  brain patterns from evoked responses: A tutorial on multivariate pattern
  analysis applied to time series neuroimaging data.
\newblock \textit{Journal of Cognitive Neuroscience}, \textbf{29}, 677--697.
\newblock \urlprefix\url{https://doi.org/10.1162\%2Fjocn_a_01068}.

\bibitem[{Hagler et~al.(2009)Hagler, Halgren, Martinez, Huang, Hillyard and
  Dale}]{Hagler_2009}
Hagler, D.~J., Halgren, E., Martinez, A., Huang, M., Hillyard, S.~A. and Dale,
  A.~M. (2009) Source estimates for {MEG}/{EEG} visual evoked responses
  constrained by multiple, retinotopically-mapped stimulus locations.
\newblock \textit{Human Brain Mapping}, \textbf{30}, 1290--1309.
\newblock \urlprefix\url{https://doi.org/10.1002\%2Fhbm.20597}.

\bibitem[{Hebart and Baker(2018)}]{Hebart_2018}
Hebart, M.~N. and Baker, C.~I. (2018) Deconstructing multivariate decoding for
  the study of brain function.
\newblock \textit{{NeuroImage}}, \textbf{180}, 4--18.
\newblock \urlprefix\url{https://doi.org/10.1016\%2Fj.neuroimage.2017.08.005}.

\bibitem[{Hogendoorn and Burkitt(2018)}]{Hogendoorn_2018}
Hogendoorn, H. and Burkitt, A.~N. (2018) Predictive coding of visual object
  position ahead of moving objects revealed by time-resolved {EEG} decoding.
\newblock \textit{{NeuroImage}}, \textbf{171}, 55--61.
\newblock \urlprefix\url{https://doi.org/10.1016\%2Fj.neuroimage.2017.12.063}.

\bibitem[{Intriligator and Cavanagh(2001)}]{Intriligator_2001}
Intriligator, J. and Cavanagh, P. (2001) The spatial resolution of visual
  attention.
\newblock \textit{Cognitive Psychology}, \textbf{43}, 171--216.
\newblock \urlprefix\url{https://doi.org/10.1006\%2Fcogp.2001.0755}.

\bibitem[{Ishai(2002)}]{Ishai_2002}
Ishai, A. (2002) Visual imagery of famous faces: Effects of memory and
  attention revealed by {fMRI}.
\newblock \textit{{NeuroImage}}, \textbf{17}, 1729--1741.
\newblock \urlprefix\url{https://doi.org/10.1006\%2Fnimg.2002.1330}.

\bibitem[{Jeffreys(1998)}]{jeffreys_1998}
Jeffreys, S.~H. (1998) \textit{The {Theory} of {Probability}}.
\newblock Oxford {Classic} {Texts} in the {Physical} {Sciences}. Oxford
  University Press, third edn.

\bibitem[{Kass and Raftery(1995)}]{Kass_1995}
Kass, R.~E. and Raftery, A.~E. (1995) Bayes factors.
\newblock \textit{Journal of the American Statistical Association},
  \textbf{90}, 773--795.
\newblock \urlprefix\url{https://doi.org/10.1080\%2F01621459.1995.10476572}.

\bibitem[{King and Dehaene(2014)}]{King_2014}
King, J.-R. and Dehaene, S. (2014) Characterizing the dynamics of mental
  representations: the temporal generalization method.
\newblock \textit{Trends in Cognitive Sciences}, \textbf{18}, 203--210.
\newblock \urlprefix\url{https://doi.org/10.1016\%2Fj.tics.2014.01.002}.

\bibitem[{Kleiner et~al.(2007)Kleiner, Brainard and Pelli}]{Kleiner_2007}
Kleiner, M., Brainard, D. and Pelli, D. (2007) What's new in psychtoolbox-3?
\newblock \textit{PERCEPTION}, \textbf{36}.

\bibitem[{Kosslyn et~al.(1993)Kosslyn, Alpert, Thompson, Maljkovic, Weise,
  Chabris, Hamilton, Rauch and Buonanno}]{Kosslyn_1993}
Kosslyn, S.~M., Alpert, N.~M., Thompson, W.~L., Maljkovic, V., Weise, S.~B.,
  Chabris, C.~F., Hamilton, S.~E., Rauch, S.~L. and Buonanno, F.~S. (1993)
  Visual mental imagery activates topographically organized visual cortex:
  {PET} investigations.
\newblock \textit{Journal of Cognitive Neuroscience}, \textbf{5}, 263--287.
\newblock \urlprefix\url{https://doi.org/10.1162\%2Fjocn.1993.5.3.263}.

\bibitem[{Kwon et~al.(2015)Kwon, Tadin and Knill}]{Kwon_2015}
Kwon, O.-S., Tadin, D. and Knill, D.~C. (2015) Unifying account of visual
  motion and position perception.
\newblock \textit{Proceedings of the National Academy of Sciences},
  \textbf{112}, 8142--8147.
\newblock \urlprefix\url{https://doi.org/10.1073\%2Fpnas.1500361112}.

\bibitem[{{Le Bihan} et~al.(1993){Le Bihan}, Turner, Zeffiro, Cuenod, Jezzard
  and Bonnerot}]{Le_Bihan_1993}
{Le Bihan}, D., Turner, R., Zeffiro, T.~A., Cuenod, C.~A., Jezzard, P. and
  Bonnerot, V. (1993) Activation of human primary visual cortex during visual
  recall: a magnetic resonance imaging study.
\newblock \textit{Proceedings of the National Academy of Sciences},
  \textbf{90}, 11802--11805.
\newblock \urlprefix\url{https://doi.org/10.1073\%2Fpnas.90.24.11802}.

\bibitem[{Lee et~al.(2012)Lee, Kravitz and Baker}]{Lee_2012}
Lee, S.-H., Kravitz, D.~J. and Baker, C.~I. (2012) Disentangling visual imagery
  and perception of real-world objects.
\newblock \textit{{NeuroImage}}, \textbf{59}, 4064--4073.
\newblock \urlprefix\url{https://doi.org/10.1016\%2Fj.neuroimage.2011.10.055}.

\bibitem[{Lins et~al.(1993)Lins, Picton, Berg and Scherg}]{Lins_1993}
Lins, O.~G., Picton, T.~W., Berg, P. and Scherg, M. (1993) Ocular artifacts in
  {EEG} and event-related potentials i: Scalp topography.
\newblock \textit{Brain Topography}, \textbf{6}, 51--63.
\newblock \urlprefix\url{https://doi.org/10.1007\%2Fbf01234127}.

\bibitem[{Luck(2014)}]{Luck_2014}
Luck, S.~J. (2014) \textit{An introduction to the event-related potential
  technique}.
\newblock MIT press.

\bibitem[{Mangun(1995)}]{MANGUN_1995}
Mangun, G.~R. (1995) Neural mechanisms of visual selective attention.
\newblock \textit{Psychophysiology}, \textbf{32}, 4--18.
\newblock
  \urlprefix\url{https://doi.org/10.1111\%2Fj.1469-8986.1995.tb03400.x}.

\bibitem[{Mechelli(2004)}]{Mechelli_2004}
Mechelli, A. (2004) Where bottom-up meets top-down: Neuronal interactions
  during perception and imagery.
\newblock \textit{Cerebral Cortex}, \textbf{14}, 1256--1265.
\newblock \urlprefix\url{https://doi.org/10.1093\%2Fcercor\%2Fbhh087}.

\bibitem[{Nanay(2009)}]{Nanay_2009}
Nanay, B. (2009) Perception and imagination: amodal perception as mental
  imagery.
\newblock \textit{Philosophical Studies}, \textbf{150}, 239--254.
\newblock \urlprefix\url{https://doi.org/10.1007\%2Fs11098-009-9407-5}.

\bibitem[{Oostenveld and Praamstra(2001)}]{Oostenveld_2001}
Oostenveld, R. and Praamstra, P. (2001) The five percent electrode system for
  high-resolution {EEG} and {ERP} measurements.
\newblock \textit{Clinical Neurophysiology}, \textbf{112}, 713--719.
\newblock \urlprefix\url{https://doi.org/10.1016\%2Fs1388-2457%2800%2900527-7}.

\bibitem[{Oosterhof et~al.(2016)Oosterhof, Connolly and Haxby}]{Oosterhof_2016}
Oosterhof, N.~N., Connolly, A.~C. and Haxby, J.~V. (2016) {CoSMoMVPA}:
  Multi-modal multivariate pattern analysis of neuroimaging data in
  matlab/{GNU} octave.
\newblock \textit{Frontiers in Neuroinformatics}, \textbf{10}.
\newblock \urlprefix\url{https://doi.org/10.3389\%2Ffninf.2016.00027}.

\bibitem[{Pearson(2019)}]{Pearson_2019}
Pearson, J. (2019) The human imagination: the cognitive neuroscience of visual
  mental imagery.
\newblock \textit{Nature Reviews Neuroscience}, \textbf{20}, 624--634.
\newblock \urlprefix\url{https://doi.org/10.1038\%2Fs41583-019-0202-9}.

\bibitem[{Pelli(1997)}]{Pelli_1997}
Pelli, D.~G. (1997) The {VideoToolbox} software for visual psychophysics:
  transforming numbers into movies.
\newblock \textit{Spatial Vision}, \textbf{10}, 437--442.
\newblock \urlprefix\url{https://doi.org/10.1163\%2F156856897x00366}.

\bibitem[{Posner(1980)}]{Posner_1980}
Posner, M.~I. (1980) Orienting of attention.
\newblock \textit{Quarterly Journal of Experimental Psychology}, \textbf{32},
  3--25.
\newblock \urlprefix\url{https://doi.org/10.1080\%2F00335558008248231}.

\bibitem[{Quax et~al.(2019)Quax, Dijkstra, van Staveren, Bosch and van
  Gerven}]{Quax_2019}
Quax, S.~C., Dijkstra, N., van Staveren, M.~J., Bosch, S.~E. and van Gerven,
  M.~A. (2019) Eye movements explain decodability during perception and cued
  attention in {MEG}.
\newblock \textit{{NeuroImage}}, \textbf{195}, 444--453.
\newblock \urlprefix\url{https://doi.org/10.1016\%2Fj.neuroimage.2019.03.069}.

\bibitem[{Reddy et~al.(2010)Reddy, Tsuchiya and Serre}]{Reddy_2010}
Reddy, L., Tsuchiya, N. and Serre, T. (2010) Reading the mind\textquotesingle s
  eye: Decoding category information during mental imagery.
\newblock \textit{{NeuroImage}}, \textbf{50}, 818--825.
\newblock \urlprefix\url{https://doi.org/10.1016\%2Fj.neuroimage.2009.11.084}.

\bibitem[{Robinson et~al.(2019)Robinson, Grootswagers and
  Carlson}]{Robinson_2019}
Robinson, A.~K., Grootswagers, T. and Carlson, T.~A. (2019) The influence of
  image masking on object representations during rapid serial visual
  presentation.
\newblock \textit{{NeuroImage}}, \textbf{197}, 224--231.
\newblock \urlprefix\url{https://doi.org/10.1016\%2Fj.neuroimage.2019.04.050}.

\bibitem[{Rouder et~al.(2009)Rouder, Speckman, Sun, Morey and
  Iverson}]{Rouder_2009}
Rouder, J.~N., Speckman, P.~L., Sun, D., Morey, R.~D. and Iverson, G. (2009)
  Bayesian t tests for accepting and rejecting the null hypothesis.
\newblock \textit{Psychonomic Bulletin {\&} Review}, \textbf{16}, 225--237.
\newblock \urlprefix\url{https://doi.org/10.3758\%2Fpbr.16.2.225}.

\bibitem[{Scolari et~al.(2012)Scolari, Byers and Serences}]{Scolari_2012}
Scolari, M., Byers, A. and Serences, J.~T. (2012) Optimal deployment of
  attentional gain during fine discriminations.
\newblock \textit{Journal of Neuroscience}, \textbf{32}, 7723--7733.
\newblock \urlprefix\url{https://doi.org/10.1523\%2Fjneurosci.5558-11.2012}.

\bibitem[{Smout et~al.(2019)Smout, Tang, Garrido and Mattingley}]{Smout_2019}
Smout, C.~A., Tang, M.~F., Garrido, M.~I. and Mattingley, J.~B. (2019)
  Attention promotes the neural encoding of prediction errors.
\newblock \textit{{PLOS} Biology}, \textbf{17}, e2006812.
\newblock \urlprefix\url{https://doi.org/10.1371\%2Fjournal.pbio.2006812}.

\bibitem[{Sprague and Serences(2013)}]{Sprague_2013}
Sprague, T.~C. and Serences, J.~T. (2013) Attention modulates spatial priority
  maps in the human occipital, parietal and frontal cortices.
\newblock \textit{Nature Neuroscience}, \textbf{16}, 1879--1887.
\newblock \urlprefix\url{https://doi.org/10.1038\%2Fnn.3574}.

\bibitem[{Sprague and Serences(2015)}]{Sprague_2015}
--- (2015) Using human neuroimaging to examine top-down modulation of visual
  perception.
\newblock In \textit{An Introduction to Model-Based Cognitive Neuroscience},
  245--274. Springer New York.
\newblock \urlprefix\url{https://doi.org/10.1007\%2F978-1-4939-2236-9_12}.

\bibitem[{Tang et~al.(2020)Tang, Ford, Arabzadeh, Enns, Visser and
  Mattingley}]{Tang_2020}
Tang, M.~F., Ford, L., Arabzadeh, E., Enns, J.~T., Visser, T. A.~W. and
  Mattingley, J.~B. (2020) Neural dynamics of the attentional blink revealed by
  encoding orientation selectivity during rapid visual presentation.
\newblock \textit{Nature Communications}, \textbf{11}.
\newblock \urlprefix\url{https://doi.org/10.1038\%2Fs41467-019-14107-z}.

\bibitem[{Tang et~al.(2018)Tang, Smout, Arabzadeh and Mattingley}]{Tang_2018}
Tang, M.~F., Smout, C.~A., Arabzadeh, E. and Mattingley, J.~B. (2018)
  Prediction error and repetition suppression have distinct effects on neural
  representations of visual information.
\newblock \textit{{eLife}}, \textbf{7}.
\newblock \urlprefix\url{https://doi.org/10.7554\%2Felife.33123}.

\bibitem[{Tootell et~al.(1998)Tootell, Hadjikhani, Vanduffel, Liu, Mendola,
  Sereno and Dale}]{Tootell_1998}
Tootell, R. B.~H., Hadjikhani, N.~K., Vanduffel, W., Liu, A.~K., Mendola,
  J.~D., Sereno, M.~I. and Dale, A.~M. (1998) Functional analysis of primary
  visual cortex (v1) in humans.
\newblock \textit{Proceedings of the National Academy of Sciences},
  \textbf{95}, 811--817.
\newblock \urlprefix\url{https://doi.org/10.1073\%2Fpnas.95.3.811}.

\bibitem[{Wagenmakers(2007)}]{Wagenmakers_2007}
Wagenmakers, E.-J. (2007) A practical solution to the pervasive problems ofp
  values.
\newblock \textit{Psychonomic Bulletin {\&} Review}, \textbf{14}, 779--804.
\newblock \urlprefix\url{https://doi.org/10.3758\%2Fbf03194105}.

\bibitem[{Wetzels et~al.(2011)Wetzels, Matzke, Lee, Rouder, Iverson and
  Wagenmakers}]{Wetzels_2011}
Wetzels, R., Matzke, D., Lee, M.~D., Rouder, J.~N., Iverson, G.~J. and
  Wagenmakers, E.-J. (2011) Statistical evidence in experimental psychology.
\newblock \textit{Perspectives on Psychological Science}, \textbf{6}, 291--298.
\newblock \urlprefix\url{https://doi.org/10.1177\%2F1745691611406923}.

\bibitem[{Wetzels and Wagenmakers(2012)}]{Wetzels_2012}
Wetzels, R. and Wagenmakers, E.-J. (2012) A default bayesian hypothesis test
  for correlations and partial correlations.
\newblock \textit{Psychonomic Bulletin {\&} Review}, \textbf{19}, 1057--1064.
\newblock \urlprefix\url{https://doi.org/10.3758\%2Fs13423-012-0295-x}.

\bibitem[{Worden et~al.(2000)Worden, Foxe, Wang and Simpson}]{Worden_2000}
Worden, M.~S., Foxe, J.~J., Wang, N. and Simpson, G.~V. (2000) Anticipatory
  biasing of visuospatial attention indexed by retinotopically specific
  $\upalpha$-bank electroencephalography increases over occipital cortex.
\newblock \textit{The Journal of Neuroscience}, \textbf{20}, RC63--RC63.
\newblock
  \urlprefix\url{https://doi.org/10.1523\%2Fjneurosci.20-06-j0002.2000}.

\bibitem[{Xie et~al.(2020)Xie, Kaiser and Cichy}]{Xie_2020}
Xie, S., Kaiser, D. and Cichy, R.~M. (2020) Visual imagery and perception share
  neural representations in the alpha frequency band.
\newblock \textit{Current Biology}, \textbf{30}, 2621--2627.e5.
\newblock \urlprefix\url{https://doi.org/10.1016\%2Fj.cub.2020.04.074}.

\bibitem[{Zellner and Siow(1980)}]{Zellner_1980}
Zellner, A. and Siow, A. (1980) Posterior odds ratios for selected regression
  hypotheses.
\newblock \textit{Trabajos de Estadistica Y de Investigacion Operativa},
  \textbf{31}, 585--603.
\newblock \urlprefix\url{https://doi.org/10.1007\%2Fbf02888369}.

\end{thebibliography}

\newpage

\section{Supplementary Materials}
\setcounter{figure}{1}

\renewcommand{\thesubsection}{S~\arabic{subsection}}

\renewcommand{\thefigure}{S~\arabic{subsection}}

\subsection{Analysis of correct trials only}
\label{S1}
\begin{figure}[h]
\centering
\includegraphics[width=14cm]{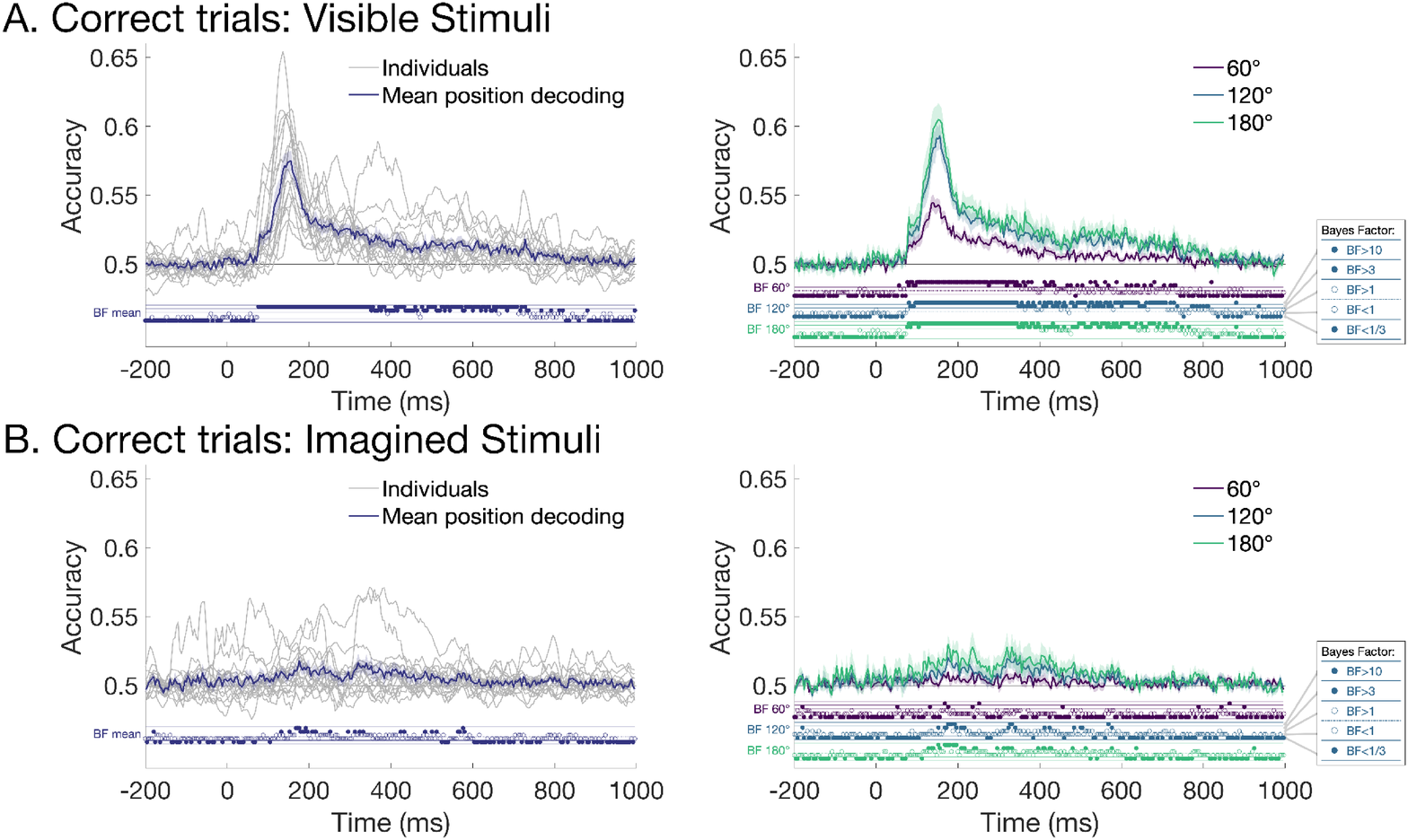}
\caption{Decoding position using correct trials only. A) Visible stimuli. B) Imagined stimuli. The results are largely the same as the original decoding analyses which used all trials regardless of performance on the task.}
\label{Figure_S1}
\end{figure}

Performance on the task was high (M > 80\%). We chose to analyse all trials in the decoding
analyses. However, it is possible that on incorrect sequences participants did not track the
stimulus correctly and would have neural responses consistent with the wrong position, affecting
the decoding. To assess position-related information on correct trials, the time-resolved decoding
analysis was performed again by excluding incorrect sequences from the test set. Number of trials
included are listed in Table \ref{Table_S1}. As can be seen in Figure \ref{Figure_S1}, the temporal dynamics of visible and
imagined stimulus position are very similar to the original set of analyses.

\begin{table}[h]

\caption{Mean numbers of stimuli per condition. Range across participants is given in brackets.}
\resizebox{\textwidth}{!}{%
\begin{tabular}{l c c c}
\headrow
                    & Pattern estimator
- Training set & Visible stimuli on
tracking task & Imagined stimuli on
tracking task \\ 
All trials          & 959.81
{[}957-960{]}             & 1432.40
{[}1394-1452{]}          & 1467.20
{[}1425-1452{]}           \\ 
\hiderowcolors
Correct trials only & As above
(passive task)          & 1240.20
{[}904-1408{]}           & 1271.80
{[}956-1446{]}
     
\end{tabular}%
}
\label{Table_S1}
\end{table}

\newpage
\subsection{Bayes Factors associated with position decoding}
\label{S2}
\begin{figure}[h]
\centering
\includegraphics[width=14cm]{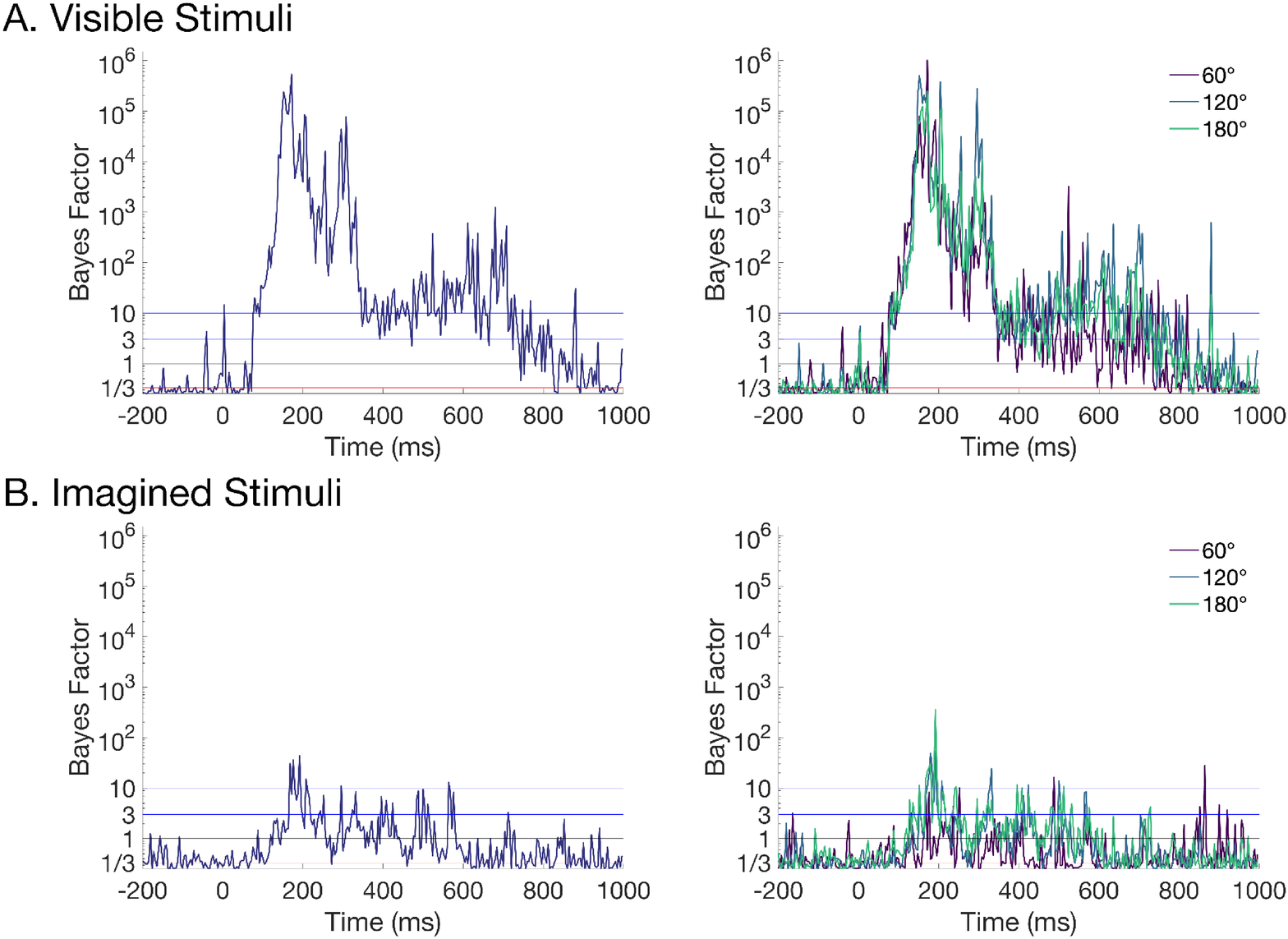}
\caption{Bayes Factors (BFs) associated with position decoding accuracy as a function of time, plotted on log scale. A) Visible stimuli. B) Imagined stimuli. Left plots show BFs associated with decoding all stimulus positions, and right plots show BFs associated with decoding according to the angular distance between position pairs.}
\label{Figure_S2}
\end{figure}

Figure \ref{Figure_S2} shows the Bayes Factors associated with whole brain position decoding from Figure \ref{Figure_4}. For both visible and imagined stimuli, there was considerable evidence for position-specific representations in the neural signal.

\newpage
\subsection{Analysis of position-related activity within frontal electrodes}
\label{S3}
To assess the contribution of potential eye movements to the decoding results (and complement
the posterior analysis), we performed decoding using a subset of electrodes from the front of the
head. The 27 electrodes were Fp1, Fp2, AFz, AF3, AF4, AF7, AF8, Fz, F1, F2, F3, F4, F5, F6, F7, F8,
FT7, FT8, FT9, FT10, FCz, FC1, FC2, FC3, FC4, FC5, FC6. Figure \ref{Figure_S3.1} shows decoding accuracy for
classifiers trained and tested on the pattern estimator. Decoding was reliably above chance from
approximately 150ms but considerably lower than whole brain analyses (Figure \ref{Figure_3}). For the crossdecoding analysis, when trained on the pattern estimator and tested on the tracking task, Figure
\ref{Figure_S3.2} shows again that decoding is lower in general than the whole brain analysis. Furthermore, for imagined stimuli there is little evidence that the frontal electrodes contained position-specific
representations (i.e., decoding is not reliably above chance except for a brief period just after
200ms).

\renewcommand{\thefigure}{S~\arabic{subsection}.{1}}
\begin{figure}[h]
\centering
\includegraphics[width=13cm]{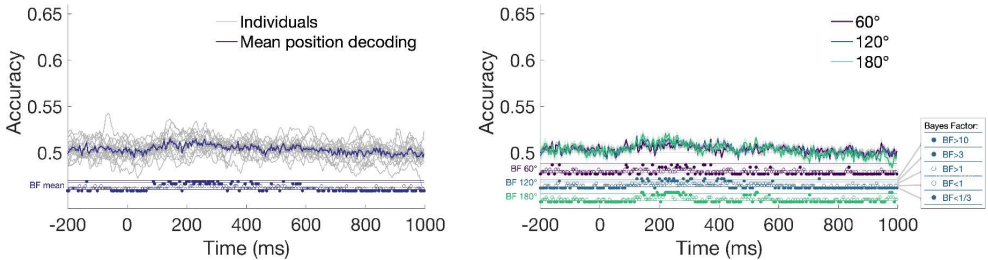}
\caption{Position decoding over time for stimuli in the pattern estimator sequences using only frontal electrodes. Left plot shows mean position decoding and right plot shows position decoding as a function of the angular distance
between stimuli.}
\label{Figure_S3.1}
\end{figure}

\renewcommand{\thefigure}{S~\arabic{subsection}.{2}}
\begin{figure}[h]
\centering
\includegraphics[width=12.5cm]{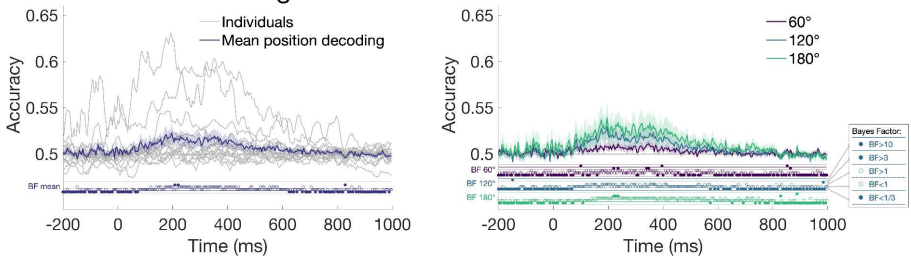}
\caption{Decoding visible and imagined stimuli on the tracking task using only frontal electrodes. A) Visible stimuli. B) Imagined stimuli.}
\label{Figure_S3.2}
\end{figure}

\newpage
\subsection{Decoding of stimuli in the tracking task by training and testing on the same condition}
\label{S4}
\renewcommand{\thefigure}{S~\arabic{subsection}}
All original analyses were performed by training classifiers on the pattern estimator sequences and
testing on the tracking task. However, the cross-decoding analysis limits the results to information
that is common to both types of experimental sequences. To assess position-specific information
in the tracking task alone, we performed cross-validated leave-one-block-out decoding separately
for the visible and imagined stimuli. Figure \ref{Figure_S4} shows that decoding accuracy was above chance for
the whole time period, including prior to the stimulus being presented or imagined. Due to the
predictable movement of the stimuli, above chance decoding prior to the stimulus could reflect
anticipatory neural signals relating to the upcoming position, and/or decoding of the previous
stimulus position.
The dynamics of the visible decoding looked qualitatively similar to, but higher than, the original
cross-decoding analysis, with a peak around 150ms. Decoding of the imagined stimuli followed a
different trajectory, with highest decoding at approximately 0ms, which was the time the tone
was presented and when participants were meant to be imagining the stimulus position.
Interestingly, imagined decoding resembled decoding in the cross-decoding time generalisation
analysis (see Figure \ref{Figure_6}B), although again with higher decoding accuracy across the whole time
period. It is impossible to make any strong claims about visible and imagined stimulus information
based on these analyses because of the confounding positions of the previous and following
stimuli. Nevertheless, it seems that the stimulus position information in the tracking task is not
drastically different to stimulus position information in the pattern estimator, lending support to
the idea that spatial imagery relies on stimulus-driven processes.
\begin{figure}[h]
\centering
\includegraphics[width=14cm]{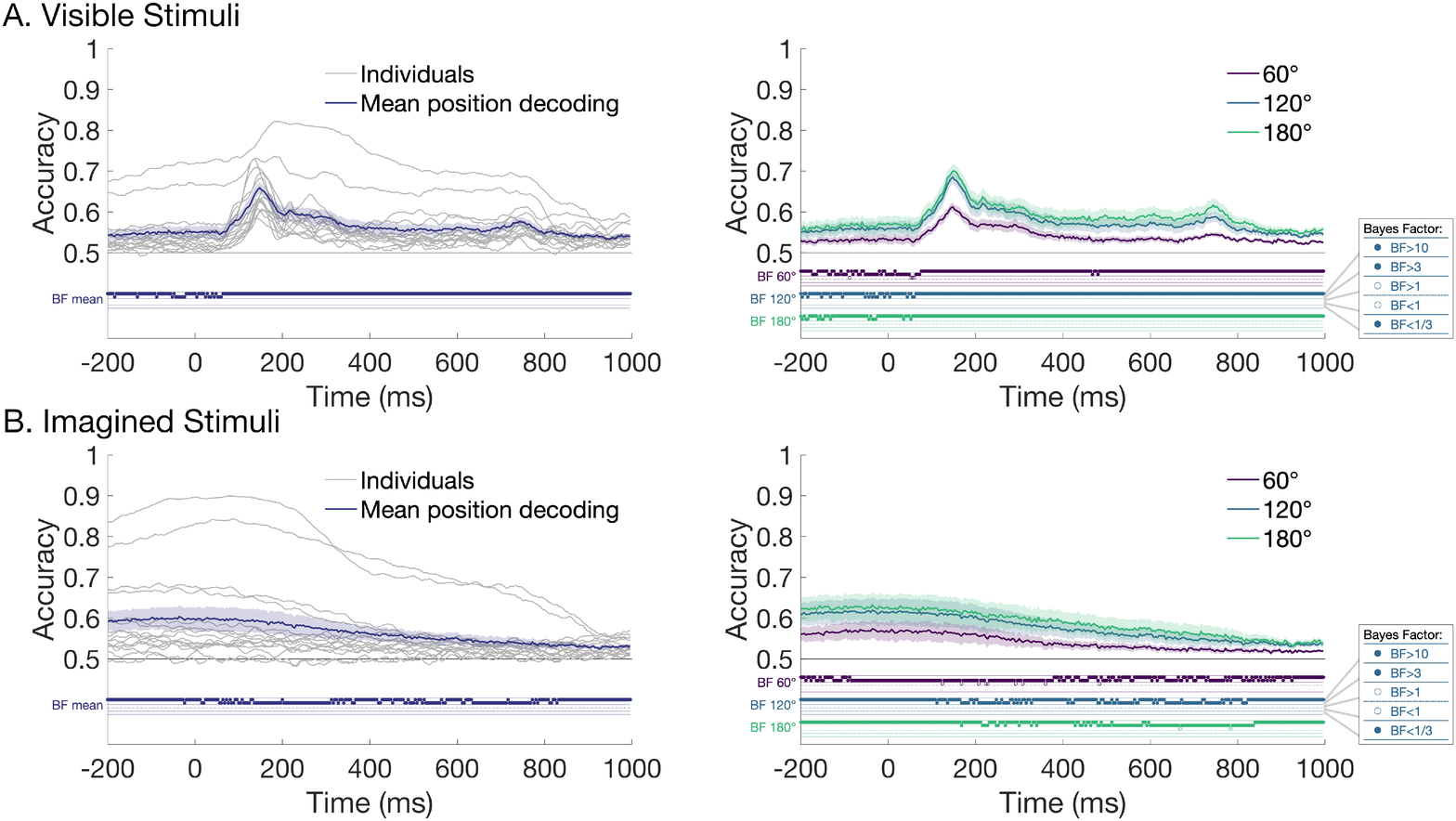}
\caption{Decoding using leave-one-block-out cross-validation for stimuli on the tracking task. A) Visible stimuli. B) Imagined stimuli.}
\label{Figure_S4}
\end{figure}

\newpage
\subsection{Neural responses to stimuli}
\label{S5}
To assess the data quality and the temporal dynamics of neural activation for the different
conditions in the study, we computed event-related potentials by averaging the signal from all
relevant trials and electrodes over time, regardless of stimulus position (Figure \ref{Figure_S5.1}). Data were preprocessed as in the decoding analysis (0.1Hz highpass, 100Hz lowpass, average reference). No
trials were rejected. ERPs were calculated separately for frontal electrodes and posterior
electrodes. For the pattern estimator, there were 960 trials per participant, except for one
participant who only had 957 trials due to a data collection error. The number of trials was similar
for visible stimuli (\emph{M} = 1432.40, range 1394-1452) and imagined stimuli (\emph{M} = 1467.20, range 1425-1475) per participant (see Table \ref{Table_S1}S1).
As can be seen in Figure \ref{Figure_S5.1}, there was a clear event-related response for all stimuli. ERPs for the pattern estimator resembled a sine wave due to the rapid stimulus presentation. Dynamics for the
slower tracking task look like characteristic event-related responses; for example, there was a
positive peak at 100ms (P1 component) and negative peak at 180ms (N170 component) after
every stimulus presentation. Imagined responses were slightly reduced relative to the visible
stimuli, though note that no statistics were performed as this was a qualitative analysis. ERPs for
each participant are shown in Figure \ref{Figure_S5.2}.

\renewcommand{\thefigure}{S~\arabic{subsection}.{1}}
\begin{figure}[h]
\centering
\includegraphics[width=14cm]{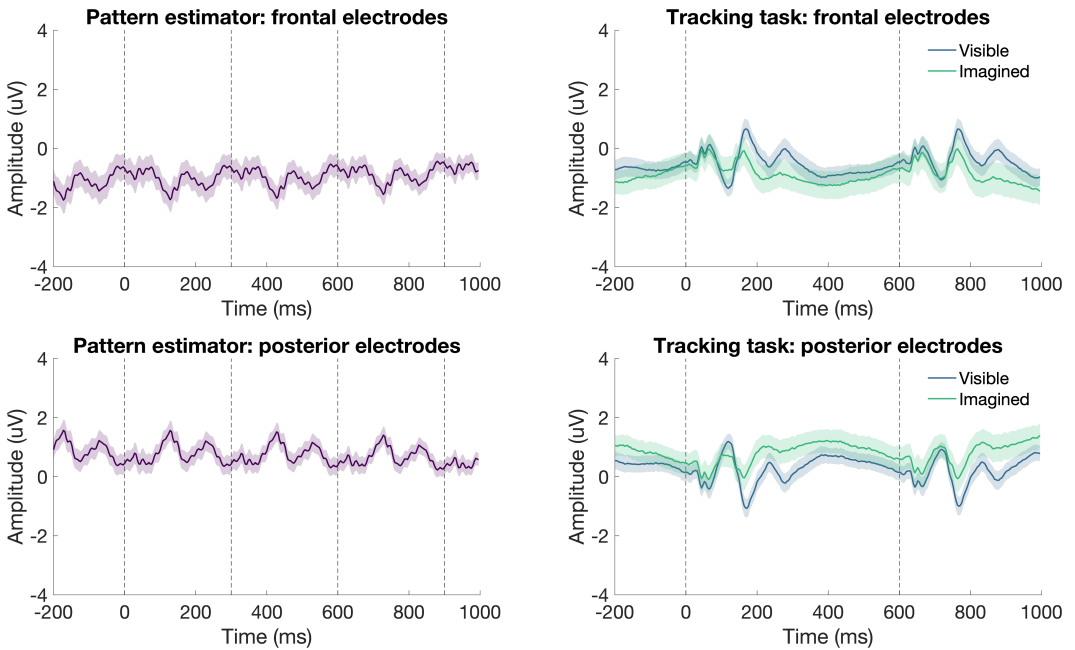}
\caption{Event-related potentials for each condition and electrode cluster. Left plots show ERPs for stimuli in the pattern estimator sequences. Right plots show ERPs for visible and imagined stimuli in the tracking task. Dotted vertical lines denote onset of stimuli within the task. Shaded areas show standard error of the mean.}
\label{Figure_S5.1}
\end{figure}

\newpage
\renewcommand{\thefigure}{S~\arabic{subsection}.{2}}
\begin{figure}[h!]
\centering
\includegraphics[width=14cm]{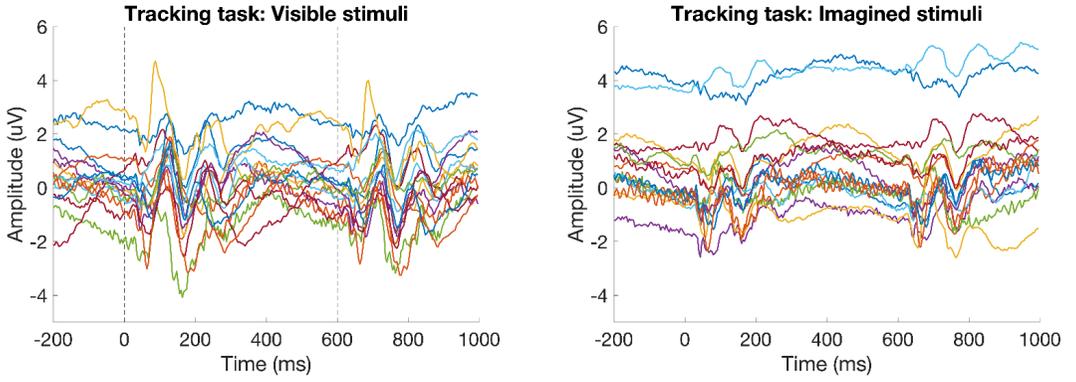}
\caption{Event-related potentials from posterior electrode cluster. Left plot shows visible ERPs and right plot shows imagined ERPs.}
\label{Figure_S5.2}
\end{figure}

\newpage
\subsection{Individual participant decoding}
\label{S6}
There was considerable variation in the magnitude of decoding accuracy across participants (see Figure S6). Imagined decoding showed more temporal variation across participants than visible decoding.
\renewcommand{\thefigure}{S~\arabic{subsection}}

\begin{figure}[h]
\centering
\includegraphics[width=14cm]{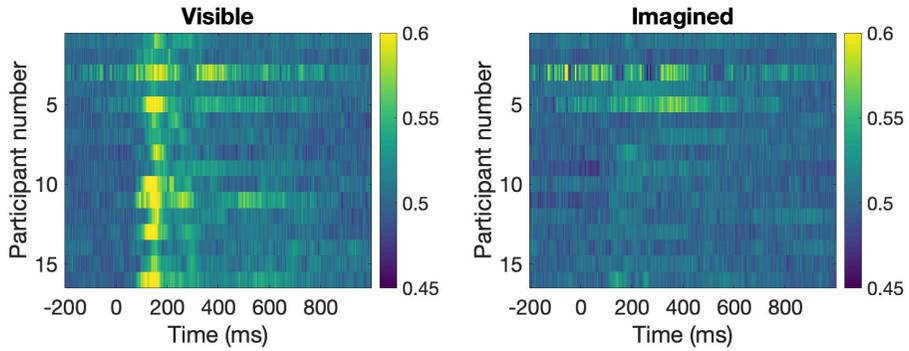}
\caption{Decoding accuracy over time per participant for visible (left) and imagined (right) stimuli on the tracking task. Decoding models were trained on the pattern estimator.}
\label{Figure_S6}
\end{figure}

\end{document}